\documentclass[twocolumn,longbib]{aastex701}
\submitjournal{PASP}

\usepackage[T1]{fontenc}
\usepackage{amsmath}
\usepackage{booktabs}
\usepackage{microtype}
\let\UrlBreaks\xxUrlBreaks
\usepackage{tikz}
\usetikzlibrary{shapes.geometric,arrows.meta,positioning,shadows,fit,patterns}
\pgfdeclarelayer{background}
\pgfsetlayers{background,main}

\definecolor{hubgreen}{HTML}{E2F0D9}
\definecolor{hubtext}{HTML}{385723}
\definecolor{actionblue}{HTML}{DDEBF7}
\definecolor{actiontext}{HTML}{1F4E78}
\definecolor{synthgold}{HTML}{FFF2CC}
\definecolor{synthtext}{HTML}{7F6000}

\newcommand{\ac}{Astro-COLIBRI}
\newcommand{\gcn}{GCN Circulars}
\newcommand{\rcab}{\ensuremath{R_{\mathrm{C}}}\mbox{-equivalent AB}}
\newcommand{\code}[1]{\texttt{\small #1}}
\newcommand{\parser}{\code{astro-colibri-circular-parser}}

\begin{document}

\shorttitle{AI-Assisted GCN Follow-up Extraction}
\shortauthors{Sch\"ussler et al.}

\title{AI-Assisted Extraction of Follow-up Observations from GCN Circulars in Astro-COLIBRI}

% Cellier, Ciric and Saint-Paul have no ORCID iD (confirmed 2026-07-30).
% AASTeX v7 requires an \email for every author. Without the [show] option the
% address is stored but not printed, so all authors share the project contact.
\author[0000-0003-1500-6571]{Fabian Sch\"ussler}
\affiliation{Université Paris-Saclay / CEA / IRFU, F-91191 Gif-sur-Yvette, France}
% \correspondingauthor must follow \email here: in AASTeX v7.0.1 the
% corresponding-author branch of \email skips the bookkeeping that the
% class later checks, so declaring it first aborts the compile.
\email[show]{astro.colibri@gmail.com}
\correspondingauthor{Fabian Sch\"ussler}
\author[0009-0005-6643-1473]{S. Bisero}
\affiliation{Université Paris-Saclay / CEA / IRFU, F-91191 Gif-sur-Yvette, France}\email{astro.colibri@gmail.com}
\author{M. Cellier}
\affiliation{Université Paris-Saclay / CEA / IRFU, F-91191 Gif-sur-Yvette, France}\affiliation{École d'ingénieurs aéronautique et spatiale (IPSA), Paris}\email{astro.colibri@gmail.com}
\author{A. Ciric}
\affiliation{Université Paris-Saclay / CEA / IRFU, F-91191 Gif-sur-Yvette, France}\email{astro.colibri@gmail.com}
\author[0009-0003-0039-0483]{B. Cornejo}
\affiliation{Université Paris-Saclay / CEA / IRFU, F-91191 Gif-sur-Yvette, France}\email{astro.colibri@gmail.com}
\author[0000-0001-5180-2845]{I. Jaroschewski}
\affiliation{Université Paris-Saclay / CEA / IRFU, F-91191 Gif-sur-Yvette, France}\email{astro.colibri@gmail.com}
\author[0000-0001-7964-4420]{A. Kaan Alkan}
\affiliation{NASA Astrophysics Data System, Smithsonian Astrophysical Observatory, Cambridge, MA 02138, USA}\email{astro.colibri@gmail.com}
\author[0000-0002-9108-5059]{W. Kiendrébéogo}
\affiliation{Université Paris-Saclay / CEA / IRFU, F-91191 Gif-sur-Yvette, France}\email{astro.colibri@gmail.com}
\author{A. Saint-Paul}
\affiliation{Université Paris-Saclay / CEA / IRFU, F-91191 Gif-sur-Yvette, France}\affiliation{École d'ingénieurs aéronautique et spatiale (IPSA), Paris}\email{astro.colibri@gmail.com}
\begin{abstract}
We present a new \ac{} component that converts free-text GCN Circulars into structured, event-linked follow-up records and combines them with reports submitted directly by the community. A continuously running listener associates new Circulars with transient events, applies deterministic pre-analysis, and invokes a schema-constrained large language model to extract photometry, contacts, redshifts, and related metadata. The records are normalized into a common event-level database and exposed through the web and mobile interfaces as report summaries, contact tools, optical-afterglow context figures, and downloadable CSV or VOTable products. All 1,775 Circulars of an operational evaluation corpus covering the first half of 2026 completed the workflow without failures. An internal human audit of 210 Circulars found 223 of 231 assessed reports (96.5\%) fully confirmed, each of the eight remaining ones containing at least one incorrect field, predominantly associated with observation timing and facility attribution. Extracted redshifts agree with three published compilations for 91.7 to 99.3\% of the events they share, every disagreement tracing to a limit, a candidate-host estimate, or a later-refined value rather than to a misread Circular. The extraction also recovers 90 to 97\% of the redshifts those compilations list for bursts it processed. Applied to the GCN archive since 2016, the pipeline yields 68,393 observations from 26,811 reports across 5,787 transient events, and now runs in real time on new Circulars. The reusable parsing and normalization pipeline is released as the open-source Python package \parser{}.
\end{abstract}


\section{Introduction}

Time-domain and multi-messenger astronomy is increasingly limited not by the number of new transient detections, but by the speed with which follow-up information can be shared, discovered, interpreted, and acted upon. Transients across the electromagnetic spectrum and all cosmic messengers, such as gamma-ray bursts (GRBs), gravitational-wave and high-energy neutrino events, X-ray transients, fast radio bursts, and supernovae, require rapid coordination across facilities with different wavelength coverage, fields of view, sensitivities, visibility constraints, and possibly proprietary-data policies. Follow-up observations identify candidate afterglows in additional wavelength ranges, confirm their association with the triggering event, measure temporal and spectral evolution, and provide spectroscopic classifications and redshifts that turn a discovery alert into a physically interpretable astrophysical source. Often, these observations must be obtained within minutes to hours because many counterpart emissions fade rapidly, yet they rely on costly and highly competitive resources such as telescope time, specialized instruments, and coordinated observing teams. Rapid and easy access to information already shared within the community therefore helps optimize this collective process: it reduces duplicated efforts, reveals missing wavelength or time coverage, supports prioritization of scarce resources, and makes it easier to contact the relevant teams while the source is still observable. At longer timescales the core operational question after a promising transient is often simple: who has observed the source, when did they observe it, how deep did they go, what did they measure, and whom should collaborators contact?

\ac{}\footnote{\url{https://astro-colibri.science}} was designed to support this real-time coordination problem by aggregating multi-messenger alerts and contextual information into a single platform for observers \citep{2021ApJS..256....5R, 2023Galax..11...22R}. Initially relying on machine-readable information obtained through real-time streams such as GCN Notices, several small-scale information-extraction pipelines have been added in recent years. These rely on regular expressions and are used mainly to identify names assigned to transient events in GCN Circulars as well as identify (re-)classifications or retractions. In general, the \ac{} transient-event pages in the frontend interface~\footnote{\url{https://astro-colibri.com}} provide a coherent interface to alert data and metadata, sky localization, visibility and observability tools, archival and catalog context, and communication channels to trigger follow-up observations. The addition described here extends this event-centric view from discovery alerts to the follow-up campaign itself. In particular, it targets the often crucial information contained in \gcn{}, a long-standing, public, rapid communication mechanism for GRB and transient follow-up reports \citep{barthelmy1998gcn,gcnCirculars}.

Until recently, extracting information from Circulars has been challenging because of their free-text, human-written format. But Circulars are scientifically valuable precisely because they are flexible and various attempts at introducing a more formalized structure have not succeeded so far. Authors continue to enjoy the freedom to report detections, non-detections, spectroscopic observations, redshifts, observing conditions, data-quality caveats, calibration details, and contact persons in natural language. This flexibility is difficult to reproduce with a rigid form, but it also makes Circulars hard to search and parse systematically. A human can quickly read one Circular, but a community following an evolving GRB afterglow may need to synthesize tens of Circulars from different observatories. The problem is especially acute for optical and near-infrared (NIR) afterglows: a report may contain absolute or relative observation times, filters, Vega or AB magnitudes, upper limits, tables with different column layouts, and prose qualifiers that determine whether a number is a detection, an upper limit, calibration-star photometry, a host-galaxy measurement, or an observation of an unrelated source.

Recent progress in natural-language processing (NLP), retrieval, and large language models (LLMs) has made it practical to extract structured information from heterogeneous astronomical prose, provided that domain-specific constraints and validation are applied. The present work builds upon earlier efforts by members of the \ac{} team to treat time-domain observation reports as a dedicated astrophysical text domain: the Time-Domain Astrophysics Corpus (\emph{TDAC}; \citealt{alkan-etal-2022-tdac}) introduced an annotated corpus of observation reports and first named-entity-recognition experiments, while the Enriched Corpus for Astrophysical Entities, Coreferences, and Relations (\emph{astroECR}; \citealt{alkan-etal-2024-enriching, alkan2024thesis}) expanded this resource with entity, coreference, astrophysical-relation, and celestial-object-normalization annotations. An early attempt to leverage general-purpose LLMs for this text domain was made by \citet{2023Galax..11...63S}, who probed zero- and few-shot extraction from observation reports with InstructGPT-3 and Flan-T5-XXL and highlighted the strong sensitivity of the results to prompt design. More recently, work on the GCN archive has demonstrated the utility of topic modeling and LLM-based information extraction for classifying Circular content and recovering physical quantities such as redshifts \citep{2026ApJS..283...30S}. Our contribution differs in scope and deployment: in addition to analyzing the historical Circular archive as an offline corpus, \ac{} applies a narrower, schema-constrained extraction to newly arriving Circulars in real time, immediately linking the result to live transient events. In addition, \ac{} merges the Circulars with user-submitted observation reports and exposes the output as operational products for observers. The \ac{} system described in this paper therefore brings these ideas into a real-time follow-up workflow: new Circulars are parsed automatically, the extracted observations are attached to live events, and the results are made immediately visible and downloadable through the platform.

The goals of the implementation are fourfold:
\begin{enumerate}
    \item build an event-level record of follow-up activity from public GCN Circulars and direct \ac{} observation reports;
    \item extract observation times and exposures, filters, magnitudes, upper limits, fluxes, redshifts, observatories, instruments, authors, collaborations, and contact points into a structured format with a well-defined schema;
    \item create GRB optical-afterglow data products that are updated in real time and place new measurements in archival context; and
    \item preserve the provenance and unaltered content of the original Circulars while building an enriched, machine-readable archive that enables reproducible offline and statistical analyses.
\end{enumerate}

\section{System Overview}

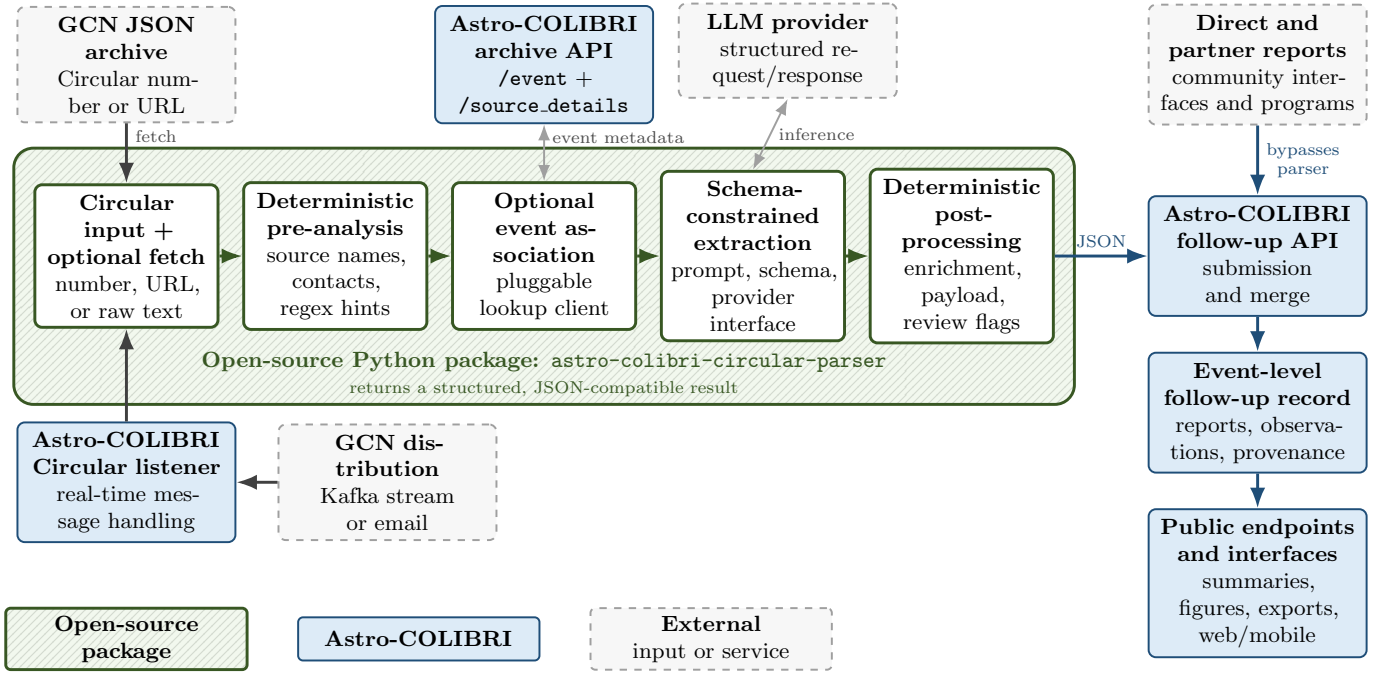
\begin{figure*}[t]
\centering
\begin{nolinenumbers}
\resizebox{\textwidth}{!}{%
\begin{tikzpicture}[
    font=\small,
    packageStep/.style={
        draw=hubtext,
        very thick,
        rounded corners=3pt,
        align=center,
        text width=2.35cm,
        minimum height=1.20cm,
        inner sep=4pt,
        fill=white
    },
    packageBoundary/.style={
        draw=hubtext,
        very thick,
        rounded corners=8pt,
        inner sep=8pt,
        % Hatching as well as colour: the green and blue fills are nearly the
        % same grey once the figure is printed in black and white.
        preaction={fill=hubgreen!45},
        pattern=north east lines,
        pattern color=hubtext!22
    },
    astroService/.style={
        draw=actiontext,
        thick,
        rounded corners=3pt,
        align=center,
        text width=2.85cm,
        minimum height=1.15cm,
        inner sep=4pt,
        fill=actionblue
    },
    externalService/.style={
        draw=gray!75,
        thick,
        dashed,
        rounded corners=3pt,
        align=center,
        text width=2.85cm,
        minimum height=1.10cm,
        inner sep=4pt,
        fill=gray!6
    },
    astroDatabase/.style={
        draw=actiontext,
        thick,
        rounded corners=3pt,
        align=center,
        text width=2.85cm,
        minimum height=1.20cm,
        inner sep=4pt,
        fill=actionblue
    },
    % The swatch carries the same hatching as the enclosure, so the legend still
    % works as a key when the figure is printed in black and white.
    legendPackage/.style={draw=hubtext, very thick, rounded corners=2pt, preaction={fill=hubgreen!45}, pattern=north east lines, pattern color=hubtext!22, align=center, text width=3.25cm, minimum height=0.62cm},
    legendAstro/.style={draw=actiontext, thick, rounded corners=2pt, fill=actionblue, align=center, text width=3.25cm, minimum height=0.62cm},
    legendExternal/.style={draw=gray!75, thick, dashed, rounded corners=2pt, fill=gray!6, align=center, text width=3.25cm, minimum height=0.62cm},
    flow/.style={-Latex, very thick, draw=black!75},
    packageFlow/.style={-Latex, very thick, draw=hubtext},
    astroFlow/.style={-Latex, very thick, draw=actiontext},
    dependency/.style={Latex-Latex, thick, draw=gray!75}
]

% Open-source package pipeline.
\node[packageStep] (packageInput) at (0,0)
{\textbf{Circular input + optional fetch}\\number, URL, or raw text};
\node[packageStep] (packageHints) at (3,0)
{\textbf{Deterministic pre-analysis}\\source names, contacts, regex hints};
\node[packageStep] (packageEvent) at (6,0)
{\textbf{Optional event association}\\pluggable lookup client};
\node[packageStep] (packageExtraction) at (9,0)
{\textbf{Schema-constrained extraction}\\prompt, schema, provider interface};
\node[packageStep] (packageOutput) at (12,0)
{\textbf{Deterministic post-processing}\\enrichment, payload, review flags};

\draw[packageFlow] (packageInput.east) -- (packageHints.west);
\draw[packageFlow] (packageHints.east) -- (packageEvent.west);
\draw[packageFlow] (packageEvent.east) -- (packageExtraction.west);
\draw[packageFlow] (packageExtraction.east) -- (packageOutput.west);

\coordinate (packageBoundaryBottom) at (6,-1.85);
\begin{pgfonlayer}{background}
\node[packageBoundary, fit=(packageInput)(packageHints)(packageEvent)(packageExtraction)(packageOutput)(packageBoundaryBottom)] (packageGroup) {};
\end{pgfonlayer}
% The label sits on the hatching, which made it hard to read in the submitted
% version; an opaque panel keeps the pattern out from behind the text.
\node[align=center, text=hubtext, fill=hubgreen!30, rounded corners=3pt,
      inner xsep=6pt, inner ysep=3pt] at (6,-1.7)
{\textbf{Open-source Python package}\\[-0.05em]
\scriptsize returns a structured, JSON-compatible result};

% Dependencies and alternative inputs.
\node[externalService] (gcnArchive) at (0,2.75)
{\textbf{GCN JSON archive}\\Circular number or URL};
\node[astroService] (eventApi) at (6,2.75)
{\textbf{\ac{} archive API}\\\code{/event} + \code{/source\_details}};
\node[externalService] (llmProvider) at (9.5,2.95)
{\textbf{LLM provider}\\structured request/response};
\node[externalService] (directReports) at (16.25,2.75)
{\textbf{Direct and partner reports}\\community interfaces and programs};

\draw[flow] (gcnArchive.south) -- node[right, pos=0.22, font=\scriptsize, text=black!70] {fetch} (packageInput.north);
\draw[dependency] (eventApi.south) -- node[right, pos=0.22, font=\scriptsize, text=black!70] {event metadata} (packageEvent.north);
\draw[dependency] (llmProvider.south) -- node[right, font=\scriptsize, text=black!70] {inference} (packageExtraction.north);

% Astro-COLIBRI real-time intake and downstream services.
\node[externalService] (gcnStream) at (3.75,-3.25)
{\textbf{GCN distribution}\\Kafka stream or email};
\node[astroService] (listener) at (0,-3.25)
{\textbf{\ac{} Circular listener}\\real-time message handling};
\draw[flow] (gcnStream.west) -- (listener.east);
\draw[flow] (listener.north) -- (packageInput.south);

\node[astroService] (followupApi) at (16.25,0)
{\textbf{\ac{} follow-up API}\\submission and merge};
\node[astroDatabase] (eventRecord) at (16.25,-2.25)
{\textbf{Event-level follow-up record}\\reports, observations, provenance};
\node[astroService] (products) at (16.25,-4.7)
{\textbf{Public endpoints and interfaces}\\summaries, figures, exports, web/mobile};

\draw[astroFlow] (packageOutput.east) -- node[above, font=\scriptsize, text=actiontext] {JSON} (followupApi.west);
\draw[astroFlow] (directReports.south) -- node[right, align=center, font=\scriptsize, text=actiontext] {bypasses\\parser} (followupApi.north);
\draw[astroFlow] (followupApi.south) -- (eventRecord.north);
\draw[astroFlow] (eventRecord.south) -- (products.north);

% Redundant visual encoding keeps the ownership distinction legible in print.
\node[legendPackage] at (0,-5.5) {\textbf{Open-source package}};
\node[legendAstro] at (4.20,-5.5) {\textbf{\ac{}}};
% The brace group after the line break is deliberate: without it the source
% reads as a backslash-backslash followed by the word input, and the
% journal's resource scanner matches the TeX file-inclusion primitive
% inside it and demands a file called "or service". The braces are a
% no-op for TeX and the box renders identically.
\node[legendExternal] at (8.40,-5.5) {\textbf{External}\\{input} or service};

\end{tikzpicture}%
}
\end{nolinenumbers}
\caption{Software boundaries of the Circular-processing workflow. The hatched green enclosure contains the stages distributed in the open-source \parser{} Python package. The package accepts a Circular number, URL, or raw text; optionally retrieves the Circular and event context; performs deterministic pre-analysis, schema-constrained extraction, enrichment, payload construction, and consistency checks; and returns a JSON-compatible result. Blue boxes denote services that remain part of the deployed \ac{} platform, while dashed gray boxes denote external inputs or services. In the real-time deployment, the \ac{} listener supplies Circular text and metadata to the package and submits its output to the follow-up ingestion service. Direct community and partner reports bypass the parser and enter the same downstream event-level record.}%
\label{fig:parser-software-boundary}
\end{figure*}

Figure~\ref{fig:parser-software-boundary} summarizes the implemented Circular-processing workflow and its software boundaries. The main real-time entry point is the \ac{} \code{circular listener}, which monitors GCN Circular messages and can also process archival Circular files. Circulars are streamed primarily through the Kafka-based distribution, with email ingestion used as a fallback. The listener consumes the version of a Circular that is distributed when it first appears. Where a Circular is subsequently revised, for example to correct a reported magnitude, the revision is not currently re-ingested and the stored record continues to reflect the original text. For each candidate message, the listener extracts the Circular number, subject, date, sender, and plain-text body, then passes the report to the reusable parser package. The parser can retrieve Circulars, generate deterministic hints, perform optional event association, run schema-constrained extraction, enrich photometry and timing information, assemble the follow-up payload, and report consistency issues. The structured payload is then submitted to the \ac{} follow-up API, which merges the report into a per-event follow-up document, regenerates derived summaries, and, where applicable, creates optical-afterglow figures and exportable photometry tables. The same event-level record drives publicly readable report summaries and optical-afterglow figures, together with authenticated export and fitting operations. Full endpoint documentation is provided by the \ac{} API documentation \citep{APIdoc}. Direct community and partner reports bypass the parser and enter the same downstream follow-up record; these additional entry points are discussed in Section~\ref{sec:community_reports}.

\subsection{Open-source parser package}

The reusable extraction path is distributed as the installable Python package \parser{} and imported through the \code{circular\_parser} namespace. It accepts a GCN Circular number or URL, or Circular text and metadata supplied directly by a streaming listener or another application. Its high-level \code{parse\_circular} interface coordinates retrieval, deterministic hint generation, optional event association, schema-constrained LLM extraction, photometric and temporal enrichment, follow-up-payload construction, and consistency checks. The result is a plain, JSON-compatible object containing the resolved source and event context, the original Circular metadata, the used regex hints, the structured model response, the enriched follow-up payload, and any issues identified for human review. The library itself performs no database writes, API submissions, or notifications; persistence and the user-facing products remain responsibilities of the surrounding \ac{} services, as shown in Figure~\ref{fig:parser-software-boundary}.

The source tree follows these processing stages rather than the deployed service architecture. Separate modules cover GCN retrieval (\code{fetch.py}), deterministic pre-analysis (\code{hints.py}), pluggable event lookup (\code{events.py}), the extraction schema and LLM interface (\code{extraction.py}), deterministic time and photometric transformations (\code{photometry.py}), payload assembly (\code{payload.py}), and post-extraction review flags (\code{consistency.py}). The lightweight orchestration layer is contained in \code{pipeline.py}, with runtime configuration in \code{settings.py}; the same pipeline is available through the Python API and the module-based command-line interface \code{python -m circular\_parser}. A cached structured extraction and example notebook allow the deterministic downstream stages to be inspected without an LLM request, and the accompanying tests run without network access or provider credentials.

\section{Circular Ingestion and Event Association}

The \ac{} Circular listener applies the same metadata parsing in real-time and archival modes: streamed messages and files on disk are both reduced to a Circular identifier, subject, date, sender, and plain-text body before they enter the parsing package. Before an LLM call is made, the system attempts to resolve the Circular to an \ac{} event through the public \code{/event}\footnote{\url{https://astro-colibri.science/apidoc\#endpoint-event}} endpoint accessing the \ac{} database. Candidate source names are extracted from the subject and body with regular expressions for common transient naming conventions, including GRB names, mission-specific trigger identifiers, gravitational-wave candidates, and common aliases introduced in Circular prose. The resolver first tests trigger identifiers embedded in each candidate, because they are usually less ambiguous than free-text source names, and otherwise uses source-name matching with common spacing and suffix variations. If no explicit source candidate resolves, trigger identifiers extracted from the complete subject and body provide a final lookup route. If several database records refer to the same reported astrophysical source, the system preserves the association with the relevant alert and refinement records while selecting a primary record for display and downstream products. These linked records represent multiple alerts or aliases for one source; they do not represent mutually exclusive candidate counterparts, which remain a limitation discussed in Section~\ref{sec:single_source_limitation}.

If no event can be matched, the parser returns the unresolved source candidates and corresponding review flags rather than silently discarding the Circular. In the deployed \ac{} workflow, the surrounding listener nevertheless completes the extraction, submits the report under a provisional source name of the form \code{GCN 12345}, and notifies administrators. This behavior is important when the GCN Circular itself announces a transient that was not previously reported elsewhere, for example through a GCN Notice. For such Circulars, new event records are currently created manually in the \ac{} platform.

\section{Hybrid NLP Extraction}

\subsection{Regex pre-analysis}

Following extensive testing and development, we settled on an extraction strategy that is deliberately hybrid. Deterministic regular expressions are used first to identify advisory hints, not to make final scientific decisions. The pre-analysis searches for optical/NIR terms, magnitude-like expressions, redshift mentions, email addresses, source names, and context snippets around relevant phrases. These hints are serialized into the LLM prompt as ``advisory only'' information that must be verified against the Circular text. This gives the model useful anchors while preserving the requirement that extracted values must come from the Circular itself.

The same deterministic layer handles several tasks that are safer outside the LLM. Header contacts are extracted from the Circular metadata, email addresses are normalized, duplicate contacts are merged, and candidate source names and trigger aliases are generated for event lookup. The system also preserves the complete original Circular text. If a message body lacks standard GCN headers, the report text passed downstream is augmented with the available title, number, subject, date, and sender metadata. This design should make it straightforward to extend the system to other types of observation reports in the future.

\subsection{Schema-constrained LLM calls}

The core NLP step uses a pluggable LLM interface that submits a single prompt constrained by a strict JSON schema. The released package includes an implementation based on the OpenAI Responses API, while callers can supply another provider that implements the same extraction interface. The prompt contains three components: the Circular number and subject, the deterministic regex hints, and the Circular text. Operational safeguards include request timeouts, retries after failures, and an optional fallback model used if the configured primary model fails, does not satisfy the schema, or returns an unambiguously implausible value. The extraction result records the model family used, allowing later auditing of model-dependent behavior.

\begin{table*}[!t]
\centering
\caption{Main fields in the schema-constrained Circular extraction response.}
\label{tab:schema}
\small
\begin{tabular}{ll}
\toprule
\textbf{Schema component} & \textbf{Purpose} \\
\midrule
\code{event\_names} & Source names or aliases explicitly identified in the Circular. \\
\code{circular\_summary} & Report-level metadata, results, redshift fields, and summary text. \\
\code{contacts} & Contact persons extracted from the header, body, and model output. \\
\code{has\_optical\_followup} & Boolean marker separating optical/NIR follow-up from other reports. \\
\code{observations} & One entry per performed observation or per filter/magnitude pair. \\
\code{warnings} & Parser/model warnings retained for audit and debugging. \\
\bottomrule
\end{tabular}
\end{table*}

The currently deployed configuration uses the smaller ``mini''-tier \code{gpt-5.4-mini} model with low reasoning effort as the primary model and the larger \code{gpt-5.5} model as a fallback. Transient request or response failures trigger up to two additional attempts with each model before escalation, whereas an unambiguously wrong result like implausible absolute-time fields triggers the fallback directly because repeating the same semantic error with the primary model is unlikely to help. This model sequence is a deployment choice rather than a fixed property of the parser: the primary and fallback model identifiers, reasoning effort, request timeout, and retry count are runtime configuration parameters and can be changed without modifying the parser code. Both models are proprietary and are subject to retirement and to silent revision by the provider; the exact model identifiers are therefore recorded with every extraction, and any reproduction of the numbers reported here should pin the same identifiers. Because Circulars are public documents, transmitting their text to an external inference provider raises no data-protection concern, but deployments bound by stricter policies can substitute a locally hosted model through the same extraction interface.

The strict schema is central to the design. The model must return a top-level object containing event names, a Circular summary, contacts, a Boolean \code{has\_optical\_followup}, observations, and warnings. Here, the plural event-name field is intended to hold aliases or identifiers for the same astrophysical source, not alternative physical counterparts. Additional properties are rejected. This turns an otherwise free-form NLP step into a typed interface between the parser and the downstream payload.

Table~\ref{tab:schema} summarizes the main fields. The Circular summary classifies the report as a high-energy detection, optical follow-up, NIR follow-up, X-ray follow-up, radio follow-up, redshift report, spectroscopy, correction, retraction, general report, or no relevant observation. Dedicated fields for neutrino or gravitational-wave counterpart searches are not part of the current schema. It also stores the observatory, short name, instrument, role, detection and redshift flags, authors, collaborations, reported results, numeric redshift, and the excerpt supporting the redshift. Contacts include name, email, role, affiliation, and source. Observation rows store one measurement or limit per entry, including absolute and relative times, elapsed duration, exposure, filter, magnitude, uncertainty, flux fields, calibration information, source excerpt, notes, and spectrum flags. Exposure is represented by its original notation together with the number of integrations, per-integration duration, and normalized total exposure in seconds when these values are available.

The prompt contains domain-specific instructions that address common failure modes in astronomical Circulars. Planned observations are explicitly excluded: only observations that have already been performed may appear in the \code{observations} array. The model is instructed not to extract comparison-star, calibration-star, or host-galaxy photometry as afterglow photometry. Multi-filter reports must be split into one entry per filter and magnitude or limit pair. Table-like Circular sections must be parsed by column semantics rather than by position alone, with separate handling of magnitude, uncertainty, exposure, and upper-limit columns. Repeated exposures such as \code{3 x 300 s} are normalized to a count of three, a per-frame duration of 300~s, and a total exposure of 900~s while retaining the original notation. Integrated exposure is kept distinct from an explicitly reported elapsed start-to-stop duration, since detector readout and other gaps may be present. Non-optical flux measurements in X-ray, gamma-ray, or radio units are allowed as flux entries but are not treated as optical magnitudes. Each observation must retain a \code{source\_excerpt} so users can audit the value against the original text.

Table~\ref{tab:schema} necessarily condenses a considerably more detailed specification. Because the prompt and the schema together define the extraction behaviour, and because a summary of either is insufficient to reproduce it, both are distributed verbatim in the released package rather than paraphrased here: the instruction text and the complete property definitions are contained in the \code{extraction.py} module of \parser{}, at the release cited in Section~\ref{sec:software_availability}.

\section{Payload Construction and Quality Control}\label{sec:payload-quality-control}

After a successful LLM response, the package combines the extraction with event metadata. The constructed payload includes event coordinates, event time, redshift, extinction value, linked events, report metadata, full text, contacts, flags, warnings, and enriched observations. The surrounding \ac{} service, rather than the package, persists this payload.

The schema deliberately retains related metadata at both report and observation levels. Observatory information, for example, is represented through several fields, including \code{observatory}, \code{observatory\_short\_name}, \code{telescope}, and \code{instrument}; summary values are assigned at report level, where applicable, while corresponding values are also stored separately for each observation. A Circular containing several observations therefore contributes repeated field instances rather than a single value for each schema field. Time metadata is likewise represented separately for each observation through start, midpoint, and end (stop) fields.

The enrichment step computes derived quantities needed for plotting and export. For optical photometry, the first application of the pipeline, these include normalized filter information, parsed absolute observation times, time since trigger, structured exposure fields, raw magnitude fields, and a common \rcab{} magnitude used for the afterglow figure (see Section~\ref{sec:afterglow_context_figure}). The midpoint of the observation is used as the plotting time whenever available. An explicitly reported midpoint, whether absolute, MJD, or relative to the event trigger, has the highest priority. In its absence, the enrichment stage uses, in order, the midpoint of a valid start--end interval, an explicitly reported elapsed duration applied to a boundary, and a start time plus half of a usable total optical exposure. Exposure-based midpoint estimation is restricted to positive totals shorter than 12~h because unusually large values are more likely to represent a parsing or association error. If none of these sources is available, the start time is retained as an approximate midpoint when possible; an end time is used only when it is the sole available boundary, rather than dropping an otherwise usable photometric point. Conflicts among reported representations and every approximate fallback remain attached to the observation for review.

The normalization layer also harmonizes selected observatory and instrument names so that equivalent names, acronyms, and long-form labels are presented consistently. Contacts from the Circular header, body email addresses, and the LLM output are deduplicated so that the \ac{} frontend can build useful email-recipient lists without repeated addresses. Contact normalization also separates affiliations embedded in person names and decodes common anti-harvesting forms such as \code{name AT host DOT org}.

Automatic quality control is implemented in the form of explicit consistency checks before submission of the extracted information to the API for storage. These checks flag for example observation rows that are classified as detections, upper limits, or non-detections but lack a numeric magnitude; rows with a magnitude but missing or unparsed observation time; conflicting or inconsistent time representations; implausible absolute MJD values; rows with negative time since trigger; and optical/NIR rows whose filter cannot be normalized to the supported filter metadata. Radio, high-energy, and multi-messenger measurements are exempted from magnitude and optical-filter checks. Depending on the listener configuration, consistency issues and parsing failures can generate internal alerts to \ac{} developers.

The package extracts report-level redshift evidence through the schema-constrained model. When a redshift is found, the follow-up payload contains the numeric value, reported-result lines, and supporting excerpt, allowing redshift-only and photometric Circulars to appear in the same event-level overview. Separately, the deployed \ac{} platform runs a deterministic redshift-monitoring pipeline for event-metadata updates that is not part of the released package. Because redshift is a well-defined scalar quantity that changes the scientific context of an event page, the metadata path uses constrained regular expressions around redshift terminology, numeric values, source names, confirmation language, and citation context. Candidate values are consolidated across Circulars and accepted for event-information updates only when they satisfy high-reliability criteria. LLM-extracted redshifts are thus retained as contextual report evidence and compared with the registered event redshift. Agreements increase confidence, whereas discrepancies are surfaced for review rather than automatically overwriting event metadata. An event therefore carries a single registered redshift together with its provenance, while every value a Circular reported remains attached to its own report with the supporting excerpt. When the deterministic criteria are not met, for instance when a Circular gives inconsistent values or the confirmation language is tentative, the pipeline notifies the \ac{} administrators by email instead of updating the event. This favours a slower but reliable path, which is appropriate here: a redshift is reported for only a minority of GRBs, and typically days after the burst, so the operational cost of waiting for a human decision is low compared with the cost of propagating a potentially wrong value to every user of the event page.

\section{Tuning, Human Verification and Operational Deployment}\label{sec:human_verification}

Operational completion alone does not establish that the structured records faithfully reproduce the scientific content of the original Circulars. We therefore evaluated the final production parser through direct human verification. This section summarizes the preceding tuning, the human-review procedure and results, and the subsequent deployment of the validated pipeline across the \ac{} platform.

The regex layer, prompt instructions, and consistency checks were tuned using the 4,537 GCN Circulars published in 2025. Pipeline failures were reviewed and used to refine the deterministic pre-analysis, prompt, or schema instructions. After this iterative development process, all 4,537 Circulars completed the operational parsing workflow without failures. Manual verification of $\sim$30 randomly selected Circulars allowed further fine-tuning of the regex guidance and the NLP prompts.

After this tuning phase, the updated pipeline was applied to an operational evaluation corpus of 1,775 Circulars received between 2026 January 1 and June 30. All 1,775 Circulars completed retrieval, schema-constrained extraction, deterministic enrichment, automated consistency verification, payload construction, and API persistence, yielding zero operational failures. Here, an operational failure means that a Circular did not reach a persisted, verified payload after the configured retries. These statistics measure workflow completion; they do not establish that every extracted field was scientifically correct.

A multiple-reviewer campaign was started through a dedicated graphical human-verification interface. Sampling concentrated on the most recent Circulars, especially those from 2026, while also including random selections from earlier years. For each selected Circular, reviewers compared the persisted original text directly with the stored report metadata, contacts, classification, and observation fields. They assigned an overall verdict of confirmed, partly correct, or rejected and field-level verdicts of confirmed, rejected, uncertain, or not applicable; rejected fields required a correction or explanatory note, while source-supported information omitted by the extraction was recorded separately. The interface neither reran the parser nor modified the production follow-up data: each review was stored centrally under its Circular and reviewer, together with the extraction hash and an append-only audit record. Reviews were keyed separately for each reviewer, allowing selected Circulars to be assessed independently by multiple people.

Two properties of this design bound what the resulting numbers can demonstrate, and we state them explicitly. First, all reviewers are co-authors and developers of the system, and approximately 80\% of the assessments were contributed by the first author, so the campaign is an internal audit rather than an independent evaluation and is exposed to confirmation bias. Second, the protocol asks reviewers to check what the parser produced against the source text. It therefore measures the correctness of extracted values but only opportunistically detects information that the extraction omitted entirely. The results below consequently quantify precision. Recall is bounded separately, and only for the redshift field, in Section~\ref{sec:redshift_crosscheck}. Full recall for all fields would require knowing what each Circular ought to have yielded, which presupposes a reference annotation produced independently of the parser; no such resource exists for this task, as discussed in Section~\ref{sec:future_work}. The protocol also scores fidelity to the Circular as published: a value transcribed correctly from a Circular that was later corrected or retracted counts as correct here.

\begin{figure*}[t]
\centering
\includegraphics[width=\textwidth]{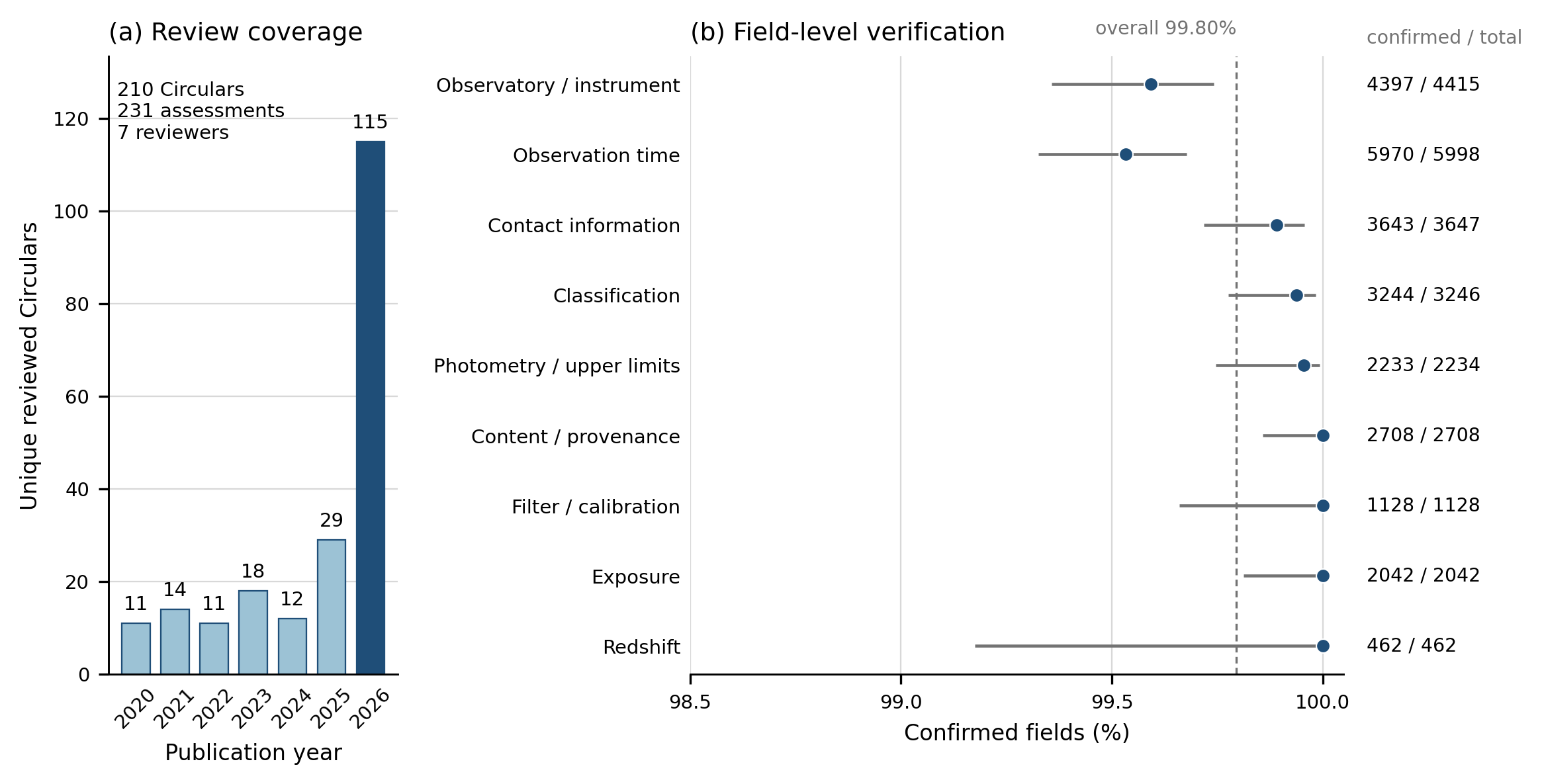}
\caption{Human verification of the final production parser using reviews submitted from 2026 July 13 onward (snapshot: 2026 July 30). (a) Publication-year distribution of the 210 unique Circulars covered by 231 assessments from seven reviewers. Twenty Circulars were independently assessed by multiple reviewers: 19 by two reviewers and one by three reviewers. (b) Confirmed fraction for definite (confirmed or rejected) field decisions, grouped by scientific content. Points show the observed fractions, horizontal bars show 95\% Wilson score intervals, and right-hand labels give the confirmed and definite counts. The dashed line marks the overall field-confirmation fraction of 25,827/25,880 (99.80\%). The seven uncertain verdicts and one source-supported missing observation are not included in the plotted fractions.}
\label{fig:human-verification}
\end{figure*}

The final evaluation comprised 231 completed assessments by seven reviewers of 210 distinct Circulars published between January 2020 and July 2026. Of these, 115 Circulars were published in 2026 and 95 were random selections from 2020--2025. Across these assessments, 223 reports (96.5\%) were fully confirmed, eight (3.5\%) were partly correct, and none were rejected. Reviewers assigned 25,887 field-level verdicts: 25,827 confirmed, 53 rejected, and seven uncertain. Excluding the uncertain verdicts, 99.80\% of the 25,880 definite field decisions were confirmed; reviewers identified only one entirely missing observation. Figure~\ref{fig:human-verification} shows the temporal coverage of the reviewed sample and the confirmation rates for grouped field categories.

The field-level fraction should not be read as the outcome of 25,880 independent trials, for two reasons. First, the total counts field instances rather than distinct schema fields: the report- and observation-level multiplicity described in Section~\ref{sec:payload-quality-control} means that a single assessment carries between 32 and 1,143 field decisions (mean 112), and the 50 largest assessments alone supply 53\% of the total. Second, errors are strongly correlated within a report. All 53 rejected fields occur in the same eight assessments that received a ``partly correct'' overall verdict, because the same few mistakes propagate across every observation row: Circular 40471 alone accounts for 24 rejections, arising from just two errors repeated across each of its six \textit{Swift}/UVOT rows. The values of \code{observatory\_short\_name} and \code{telescope} were transposed, and the observation mid-times were derived by offsetting the reported start time rather than taking the midpoint of the reported start--stop interval. Notably, the facility itself was identified correctly: \code{observatory} and \code{instrument} were confirmed in every instance, so the schema's redundant facility fields, rather than the underlying attribution, produced the rejections. The more conservative statement is therefore that eight of 231 assessed reports (3.5\%) contained at least one incorrect field, with the field-level fraction as a secondary, non-independent measure of how much of each report was affected. The fields also differ in intrinsic difficulty, so that overall fraction does not hold uniformly across all fields. Table~\ref{tab:error-taxonomy} separates them: every rejection falls in observation timing or facility attribution, while filter, calibration, exposure and redshift fields produced none at all.

\begin{table}[!htbp]
\centering
\caption{Rejected (Rej.) and uncertain (Unc.) field verdicts by scientific content, with the number of assessments (Assess.) in which they occur. The four categories not listed (content/provenance, filter/calibration, exposure, redshift) produced no rejected or uncertain verdicts. An assessment with rejections in more than one category is counted in each, so the Assess. column does not sum to the total, which is the number of distinct assessments.}
\label{tab:error-taxonomy}
\footnotesize
\begin{tabular}{lrrr}
\toprule
\textbf{Category} & \textbf{Rej.} & \textbf{Assess.} & \textbf{Unc.} \\
\midrule
Observation time & 28 & 5 & 7 \\
Observatory / instrument & 18 & 4 & 0 \\
Contact information & 4 & 1 & 0 \\
Classification & 2 & 1 & 0 \\
Photometry / upper limits & 1 & 1 & 0 \\
\midrule
Total & 53 & 8 & 7 \\
\bottomrule
\end{tabular}
\end{table}

Table~\ref{tab:error-taxonomy} groups the rejected and uncertain verdicts by scientific content. The errors are confined to five of the nine categories. Observation timing dominates, and it is also the only category in which reviewers recorded uncertain verdicts, reflecting Circulars whose time expressions are genuinely ambiguous or even inconsistent in the source text. Observatory and instrument attribution is the second contributor. Conversely, filter and calibration information, exposure fields, redshifts, and the retained provenance content produced no rejections at all. Users combining extracted rows across reports should therefore treat observation times and facility attribution as the fields most in need of verification against the original Circular.

The 20 Circulars assessed by more than one reviewer provide a check on the reproducibility of the review protocol itself. All 20 received unanimous overall verdicts, and the 2,105 field decisions made independently by more than one reviewer agreed in every case. Because the confirmed base rate is close to unity and this subset happened to contain no rejections, a chance-corrected agreement statistic is degenerate here; the result establishes that the protocol is applied consistently, not that reviewers would agree on the rare, difficult cases.

In summary, the audit found no report whose scientific content was misrepresented as a whole, and localized the errors it did find to timing and facility attribution in a small number of reports. Together with the zero operational failures on the 2026 evaluation corpus, this supports use of the pipeline for real-time situational awareness, subject to the precision-only scope and the internal composition of the reviewer panel noted above.

The pipeline was integrated into the production \ac{} service with release v2.30.0 on 2026 June 24 \citep{astroColibriReleases}. In parallel to parsing new incoming GCN Circulars in real time, the final pipeline was applied to the full GCN Circular archive since 2016, corresponding to the beginning of the \ac{} transient archive. The archive start date, rather than any property of the Circulars, sets this boundary: the extraction attaches every report to an \ac{} event record, and no such records exist for transients that occurred before the platform began ingesting alerts. In total, over 26,000 Circulars are now represented in the \ac{} platform as structured follow-up information and are accessible through the interfaces described in the following.

\subsection{The extracted archive}\label{sec:archive_characterization}

\begin{figure*}[!tbp]
\centering
\includegraphics[width=\textwidth]{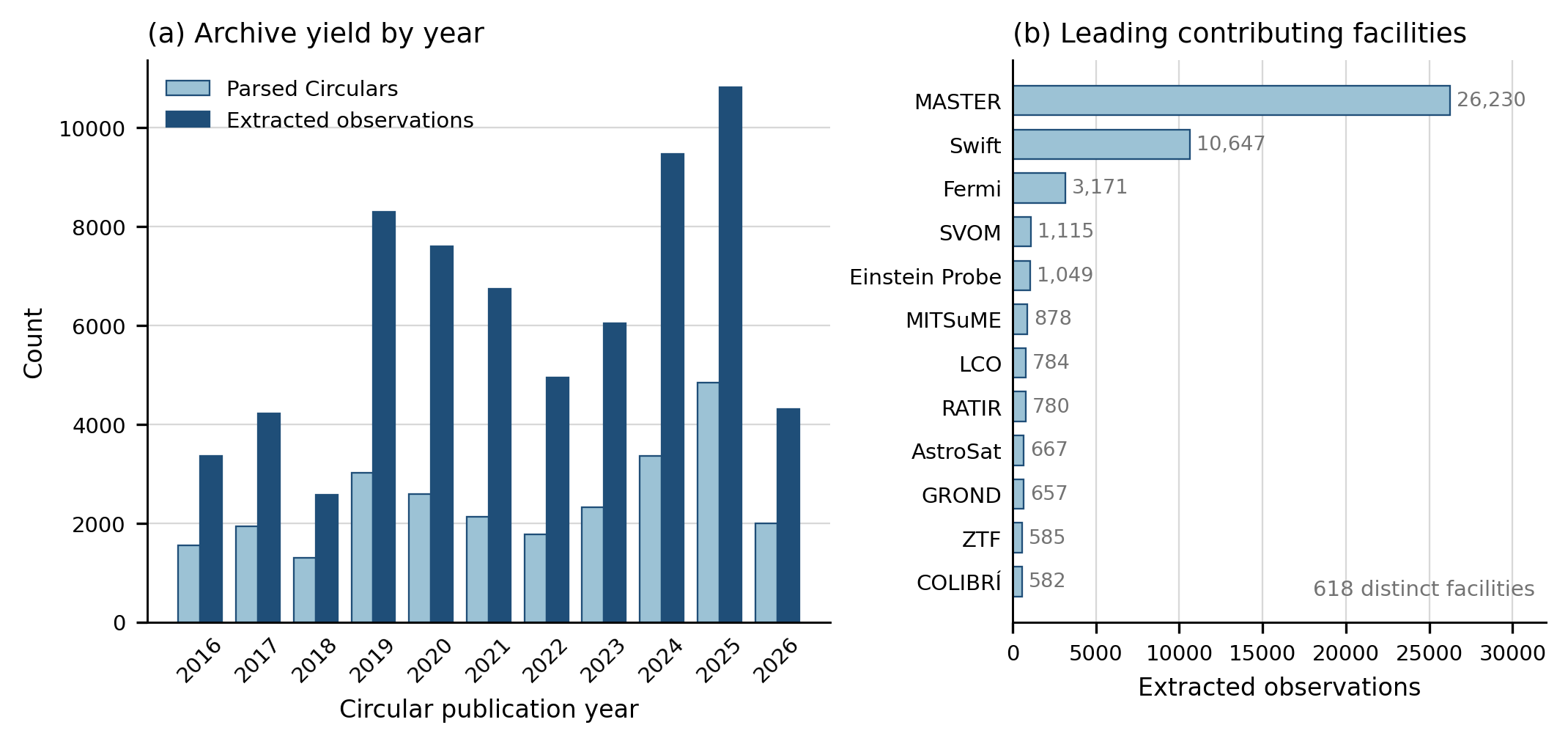}
\caption{The follow-up database produced by applying the final pipeline to the GCN Circular archive from 2016 onward (snapshot: 2026 July 30). (a) Parsed Circulars and extracted observations per Circular publication year. (b) The twelve facilities contributing the most extracted observations, out of 618 in total; instrument-level names are grouped under their facility. The wide-field MASTER network dominates because it reports many short exposures and limits per Circular.}
\label{fig:archive-extraction}
\end{figure*}

Table~\ref{tab:archive-summary} characterizes the resulting database, and Figure~\ref{fig:archive-extraction} shows its distribution over time and over contributing facilities. The archive run covers 26,811 reports attached to 5,787 transient events and yields 68,393 individual observations, of which upper limits outnumber detections by roughly two to one. This ratio is itself a useful property of the product: the non-detections that constrain a fading afterglow are recorded on the same footing as the detections, which is precisely the information an observer needs when deciding whether a target is still worth pursuing.

\begin{table}[!htbp]
\centering
\caption{The follow-up database after applying the final pipeline to the GCN Circular archive from 2016 onward (snapshot: 2026 July 30).}
\label{tab:archive-summary}
\footnotesize
\begin{tabular}{lr}
\toprule
\textbf{Quantity} & \textbf{Count} \\
\midrule
Transient events with follow-up records & 5,787 \\
\ldots{}of which with optical/NIR follow-up & 3,349 \\
Parsed reports & 26,811 \\
\ldots{}classified as optical/NIR follow-up & 12,036 \\
\ldots{}reporting photometry & 11,394 \\
Extracted observations & 68,393 \\
\ldots{}detections & 20,990 \\
\ldots{}upper limits & 42,926 \\
\ldots{}non-detections & 1,415 \\
\ldots{}without a photometric value & 3,062 \\
\ldots{}flagged as spectroscopic & 3,388 \\
Reports with a numeric redshift & 859 \\
Extracted contact records & 84,833 \\
Distinct contributing facilities & 618 \\
\bottomrule
\end{tabular}
\end{table}

Of the 50,015 observations that carry optical or NIR photometry, 47,515 (95.0\%) could be placed on the common \rcab{} scale and are therefore available to the context figures and exports; the remaining 2,500 report a filter that is not in the supported filter metadata and are retained with their raw values only. This fraction measures filter-metadata coverage over the whole archive; it is a different quantity from the field-confirmation fractions of Figure~\ref{fig:human-verification}, which record whether a reviewer confirmed an extracted photometric value, not whether its filter could be normalized. Automated quality control flagged consistency issues in 5,569 reports (20.8\%) and time-representation issues in 15,358 observations (22.5\%). These are advisory flags attached to the record for review, not failures, and their concentration in the timing fields is consistent with the human-verification results in Table~\ref{tab:error-taxonomy}. The fallback, larger LLM model was invoked for 94 of the 26,808 reports carrying a model identifier (0.35\%), so the primary, smaller model satisfied the schema and the plausibility checks in more than 99.6\% of cases. The three remaining reports are direct community submissions, which do not pass through a model at all.

\subsection{External cross-check of extracted redshifts}\label{sec:redshift_crosscheck}

The human verification of Section~\ref{sec:human_verification} compares the extraction against the Circular it came from, and therefore cannot detect an error that both the reviewer and the model would make. Redshift offers a rare opportunity for a fully independent check, because it is a scalar quantity that is also compiled from the literature by external catalogs. We therefore compared every redshift the pipeline extracted with the GRBweb compilation \citep{grbweb}, and then against two further compilations to gauge how far any single one of them can be treated as ground truth.

The pipeline recorded a numeric redshift for 469 events, of which 253 also carry a redshift in GRBweb. The cross-check and the recall estimates were evaluated on 2026 October 7. This count runs over events in the full archive and is unrelated to the redshift entry of Figure~\ref{fig:human-verification}, which counts field-level verdicts inside the 231 audited assessments. For 232 of the 253 GRBweb entries (91.7\%) at least one extracted value agrees with the catalog to within $\Delta z = 0.01$, and the median absolute difference across the matched sample is zero. We inspected all 21 non-agreeing events individually against the supporting excerpts. In every case the extracted number is present in the Circular and was read correctly; the disagreement arises from what the number means. Nine are upper or lower limits or values reported conditionally on a candidate association, for example the $z=3.87$ of a candidate host galaxy for GRB~230307A reported in \href{https://gcn.nasa.gov/circulars/33580}{GCN Circular 33580}, whose accepted redshift is 0.065. Seven are photometric redshifts of a candidate host rather than a spectroscopic redshift of the burst. The remaining five are preliminary or tentative values that were later refined, such as $z=7.21$ for GRB~250314A against the catalog value of 7.30. This accounting, however, presumes that GRBweb itself is correct.

Treating a single catalog as ground truth overstates what the comparison can settle, because the published compilations do not agree among themselves. Repeating the exercise against the GRB table of \citet{greinerTable} and the \textit{Swift} GRB table \citep{swiftGRBTable} shows that, for the bursts any two of the three share, they quote redshifts consistent to within $\Delta z=0.01$ for only 96--97\% of cases: 391 of 402 for GRBweb and Greiner, 324 of 335 for GRBweb and \textit{Swift}, and 373 of 388 for Greiner and \textit{Swift}. Against that background the extraction agrees with Greiner for 204 of 208 shared events (98.1\%) and with the \textit{Swift} table for 138 of 139 (99.3\%), both higher than the 91.7\% obtained against GRBweb alone. Four of the 21 disagreements above are in fact cases where Greiner and \textit{Swift} confirm the extracted value and GRBweb is the outlier --- two of them drawn from the candidate-host group and two from the later-refined group. One is GRB~231111A, where both tables list the $z=1.179$ we extracted and independently flag our second value, $z=1.39$, as tentative. GRB~250314A, cited above as a refined value, makes the same point more gently: GRBweb and Greiner themselves quote 7.30 and 7.27 for that burst, against the 7.21 reported in the Circular we parsed. Thirteen of the remaining events appear in neither compilation at all, which is itself consistent with our reading that those Circular values were limits or candidate-host estimates that never became accepted redshifts. The four events that remain (GRB~210104A, GRB~210321A, GRB~240619A and GRB~250314A) are listed with a secure redshift in at least one of the two additional tables that still differs from the value we extracted. In each the Circular number was read correctly but was preliminary or conditional, and the compilations quote the later refined value. The 91.7\% should therefore be read as a lower bound: part of the residual measures the scatter between reference catalogs rather than any fault of the extraction.

The comparisons above measure precision. Redshift is also the one field for which these independent catalogs support a bound on recall, because a scalar recorded externally can be looked for in our records even where we never extracted it. Restricting each catalog to the bursts for which \ac{} holds a follow-up record, so that the corresponding Circulars did pass through the pipeline, the extraction recovered a numeric redshift for 253 of the 281 events listed by GRBweb (90.0\%), 208 of 217 listed by \citet{greinerTable} (95.9\%), and 139 of 143 listed in the \textit{Swift} table (97.2\%). The ordering repeats the pattern of the precision comparison, with the weakest agreement against the most inclusive compilation. Two caveats bound how far these are recall of the extraction step itself. A catalog redshift need not have been announced in a Circular at all, since several are first published in journal papers, so the denominator includes bursts for which no Circular text existed to extract from and the recall over Circular-reported redshifts is correspondingly higher. The compilations also disagree among themselves at the few-per-cent level established above, so the denominator is itself uncertain. We therefore quote the range rather than a single number: the extraction recovers of order 90 to 97\% of externally catalogued redshifts for the bursts it processed.

Across all three references, then, no disputed value turned out to be a misreading of the Circular text, which is a reassuring result for the extraction step. The exercise does, however, identify a concrete limitation of the current schema: the report-level redshift field stores a single number without distinguishing a measurement from a limit, a candidate-host photometric redshift, or a superseded preliminary value. The supporting excerpt preserves that distinction for a human reader, and the deterministic redshift-monitoring pipeline described in Section~\ref{sec:payload-quality-control} applies its own high-reliability criteria before any event metadata is updated, so the platform does not act on these values. Making the distinction explicit in the schema is a natural next step, and could let the extracted redshifts be used directly in statistical work.

\section{Data Storage and Derived Products}\label{sec:data_products}

The \ac{} API~\citep{2023Galax..11...22R} stores parsed and user-submitted reports in a follow-up database collection. Before a new payload is merged, the service locates the target record using case-insensitive current or historical source names, the Circular identifier, and primary or linked trigger identifiers; generic event labels such as \code{retracted} are not used as event keys. A Circular classified as a retraction flags the associated event as retracted rather than removing the reports already attached to it, so the follow-up record of what was observed survives the retraction of the trigger. Corrections are narrower in practice: they most often revise the classification of an event, for example when Einstein Probe reclassifies an X-ray transient as a stellar flare, and are applied to the event (meta-) data while the underlying observation rows remain as reported. The located record is updated in place, preventing naming variants from creating duplicate campaign records. This merge step combines contacts, redshift reports, linked events, observations, report summaries, flags, figure metadata, and update timestamps. This is important because a single GRB may accumulate many Circulars over hours to weeks, while the single event page should show one coherent campaign record.

For data access, the \ac{} frontend uses publicly readable \ac{} API endpoints for report summaries (\code{/followup\_summary})\footnote{\url{https://astro-colibri.science/apidoc\#endpoint-followup_summary}} and generated figures (\code{/optical\_afterglow\_lightcurve})\footnote{\url{https://astro-colibri.science/apidoc\#endpoint-optical_afterglow_lightcurve}}, while detailed data retrieval, export creation, and model fitting require a user account. The data-product layer provides a JSON response containing trigger identifiers, source name, figure metadata, row count, observation rows, report summaries, linked events, and default fit metadata. It can also create CSV and VOTable files. The CSV export represents each observation as a row with time, MJD, time since trigger, exposure count, per-frame and total exposure, filter, raw magnitude, uncertainty, common \rcab{} magnitude, upper-limit flag, observatory, Circular identifier, and provenance fields. Each converted row also carries the spectral index $\beta$ applied in Equation~\ref{eq:color} and whether it was reported in the Circular, registered for the event, or taken as the default, so a user can reproduce or undo the color transformation and propagate its systematic independently. An additional VOTable export wraps the same data in IVOA-compatible XML, improving interoperability with a wide variety of astronomy tools~\citep{ivoaVotable}.

\subsection{Optical-Afterglow Context Figures}\label{sec:afterglow_context_figure}

\begin{figure*}[!t]
\centering
\begin{minipage}{0.56\linewidth}
\centering
\includegraphics[width=\linewidth]{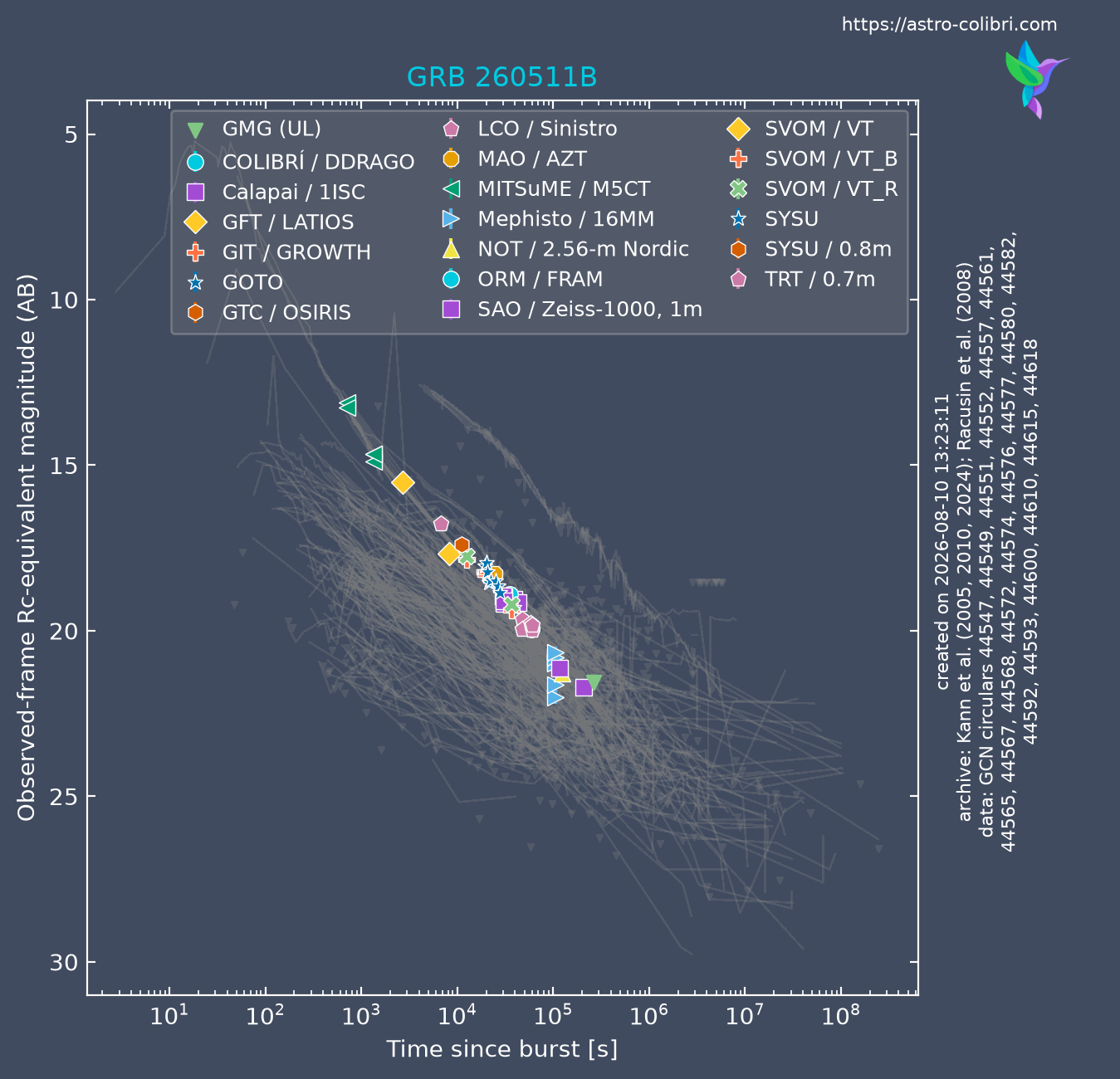}\\[-0.2em]
{\small (a) Generated optical-afterglow context figure}
\end{minipage}\hfill
\begin{minipage}{0.34\linewidth}
\centering
\includegraphics[width=\linewidth,height=0.42\textheight,keepaspectratio]{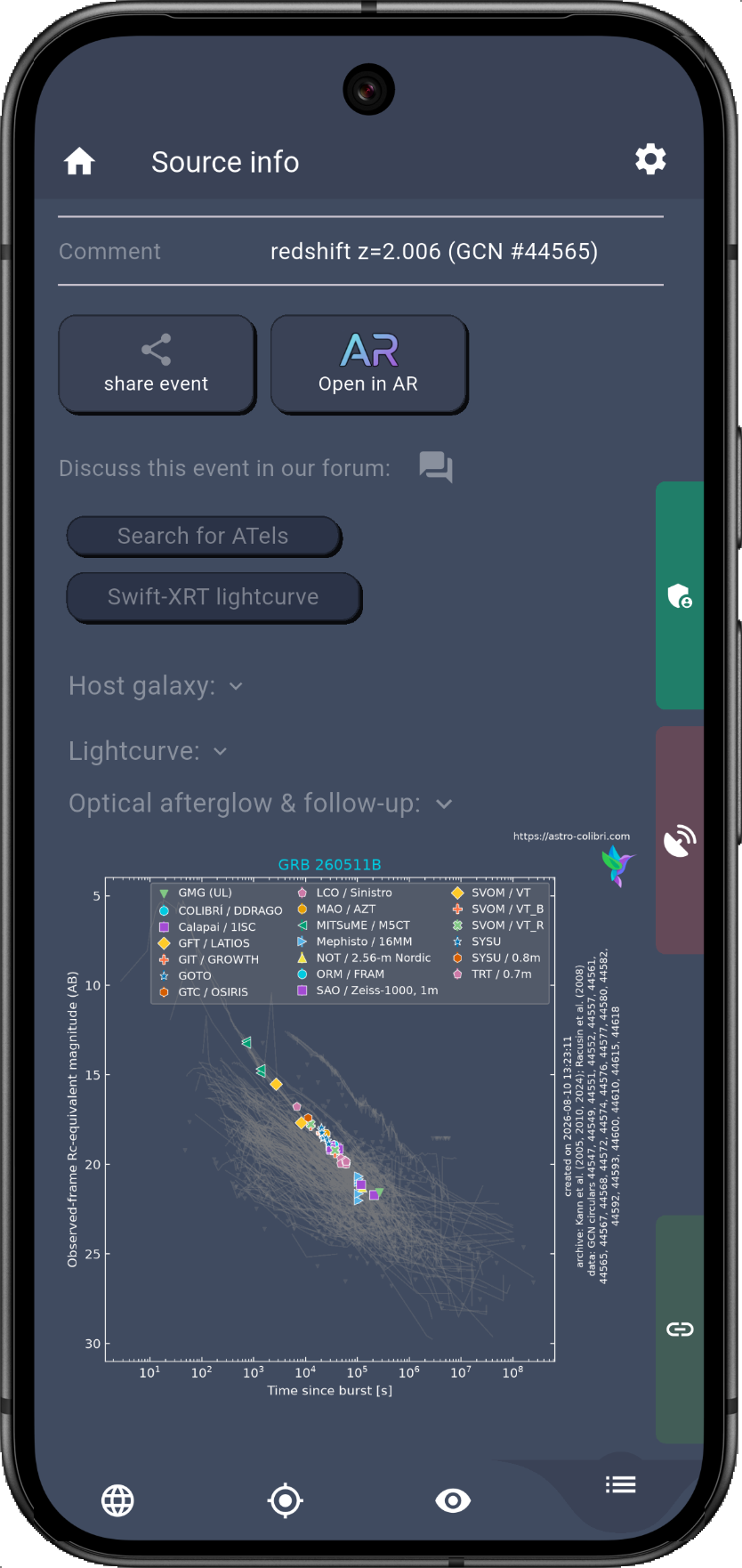}\\[-0.2em]
{\small (b) Mobile follow-up section overview}
\end{minipage}
\caption{Optical-afterglow and follow-up presentation in \ac{}. Panel (a) shows an example context figure for GRB~260511B, combining current-event photometry extracted from GCN Circulars and observation reports with archival GRB afterglow measurements on a common observer-frame \rcab{} magnitude scale. Panel (b) shows the mobile source-information panel with the generated context figure, light-curve download, fitting control, and expandable observation-report section.}
\label{fig:afterglow-example}
\end{figure*}

For GRB-like events, the API can generate an optical-afterglow context figure. The figure plots the common observer-frame \rcab{} magnitude against the time since the event trigger, using each observation's mid-time where one is available. Because one source is often detected by several instruments with slightly different trigger times, every report attached to an event is timed against a single reference: the earliest of the linked alerts, unless they are separated by more than an hour, which indicates that distinct events have been associated with the same name rather than one burst seen twice. It is intended as a rapid visual illustration of afterglow brightness and evolution, not as a publication-ready or rest-frame-corrected luminosity analysis. Figure~\ref{fig:afterglow-example} shows an example generated by the current system for GRB~260511B together with the corresponding mobile follow-up view.

The archival context is stored as a curated static table containing 8,283 measurements from 138 GRBs and XRFs. It merges the recent compilation of \citet{2024A&A...686A..56K}, distributed through VizieR~\citep{vizierKann}, with the 840 measurements in the VizieR table of \citet{2010ApJ...720.1513K}, the complete pre-\emph{Swift} TeX catalogue of \citet{KannCatalog2005}, and the machine-readable GRB~080319B supplementary light curve of \citet{2008Natur.455..183R}. The earlier catalogues add, in particular, the extensive GRB~030329 and GRB~080319B light curves that are absent from the 2024 sample. Pre-\emph{Swift} rows flagged as already corrected for Galactic extinction are excluded, because the catalogue does not record enough information to reverse the correction and return them to the observed axis used here. After conversion, exact cross-catalogue duplicates are removed, as are archival plotting outliers fainter than 30~mag, values brighter than 5~mag after $10^5$~s, and upper limits lower than 12~mag. These archive-only display cuts only prevent non-magnitude values and extreme limits from setting the figure scale; they are not applied to current-event measurements. The published tables list magnitudes in their original filters; the \rcab{} values used for plotting were derived by us with exactly the transformation described below, so the archival population and the newly extracted measurements are placed on the observer-frame comparison axis by the same procedure. The archive is used as a visual comparison population: it shows where the current event lies relative to well-studied optical afterglows, while leaving detailed physical interpretation to later analysis. For a measurement in filter $f$ with tabulated effective wavelength $\lambda_f$, a Vega-system magnitude is first placed on the AB scale:
\begin{equation}
    m_{\mathrm{AB}} = m_{\mathrm{raw}} + \Delta_{\mathrm{Vega}\rightarrow\mathrm{AB}}(f).
\end{equation}
It is then transformed to a Cousins-$R$ equivalent filter band by assuming a power-law spectral energy distribution $F_\nu \propto \nu^{-\beta}$:
\begin{eqnarray}
    m_{R_{\mathrm{C}},\mathrm{AB}} &=& m_{\mathrm{AB}} + 2.5\,\beta\,\log_{10}\left(\frac{\nu_{R_{\mathrm{C}}}}{\nu_f}\right) \\
    &=& m_{\mathrm{AB}} + 2.5\,\beta\,\log_{10}\left(\frac{\lambda_f}{\lambda_{R_{\mathrm{C}}}}\right). \label{eq:color}
\end{eqnarray}
This is a spectral color transformation to a common observer-frame band, not a redshift-dependent rest-frame K-correction; the magnitude notation follows \citet{hogg2002}. The spectral index is taken from the observation itself when a Circular reports one, otherwise from the event metadata, and otherwise from a configurable default of $\beta=0.7$ adopted as an approximate optical-afterglow plotting convention rather than an event-specific measurement. Because most Circulars report no spectral index, the default applies to the large majority of measurements, and the resulting systematic is not represented by the plotted measurement errors. Its size follows directly from Equation~(\ref{eq:color}) and grows with the distance of the filter from $R_{\mathrm{C}}$: varying $\beta$ over the range 0.5--1.0 moves a point by $\lesssim0.05$~mag in $r$, $\simeq0.1$~mag in $i$, $\simeq0.2$~mag in $g$ and $z$, and up to $\simeq0.65$~mag in $K$. Since the archival comparison population was transformed with the same convention, this systematic largely cancels between current and archival points observed in the same filter, and matters mainly when comparing across widely separated bands. Unknown filters, missing magnitude systems, or non-optical fluxes remain stored with their raw values but are not plotted at the moment.

The comparison axis is one of observed, reddened magnitudes: neither the archival table nor the newly extracted measurements are corrected for Galactic foreground extinction, so no relative offset is introduced between the two populations. Circular authors, however, occasionally report magnitudes that they have already de-reddened. Where a report is flagged as such, the enrichment stage restores the observed value by adding back the foreground term,
\begin{equation}
    m_{\mathrm{obs}} = m_{\mathrm{reported}} + \left(\frac{A_\lambda}{E(B-V)}\right)_{\!f} \, E(B-V),
\end{equation}
using the event sight-line $E(B-V)$ carried in the follow-up payload and per-filter $A_\lambda/E(B-V)$ ratios computed for $R_V=3.1$. If a report is flagged as extinction-corrected but no sight-line $E(B-V)$ is available, the value is left untouched and flagged for review rather than being placed on an inconsistent axis.

The plotting workflow reads the event-level follow-up record, selects observations with corrected optical photometry, groups archival and current measurements, draws detections and upper limits with distinct markers, annotates provenance, writes a PNG image, uploads it to a publicly accessible \ac{} storage location, and records the figure URL in the event metadata.

\section{Frontend Presentation}

\begin{figure*}[!t]
\centering
\begin{minipage}{0.49\linewidth}
\centering
\includegraphics[width=\linewidth]{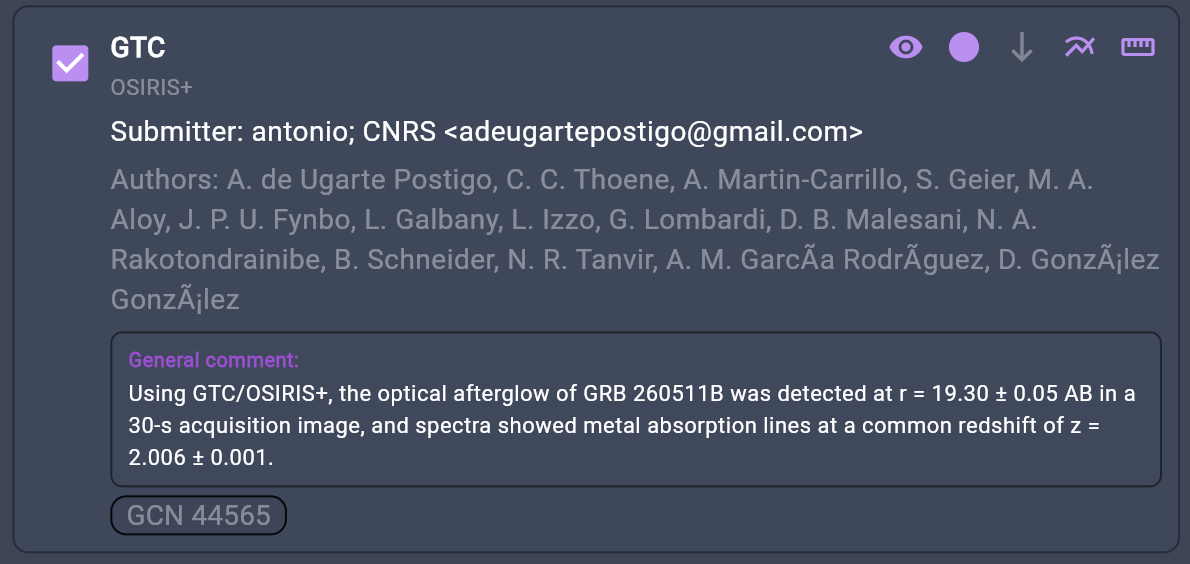}\\[-0.2em]
{\small (a) Report summary card (GCN Circular 44565)}
\end{minipage}\hfill
\begin{minipage}{0.49\linewidth}
\centering
\includegraphics[width=\linewidth]{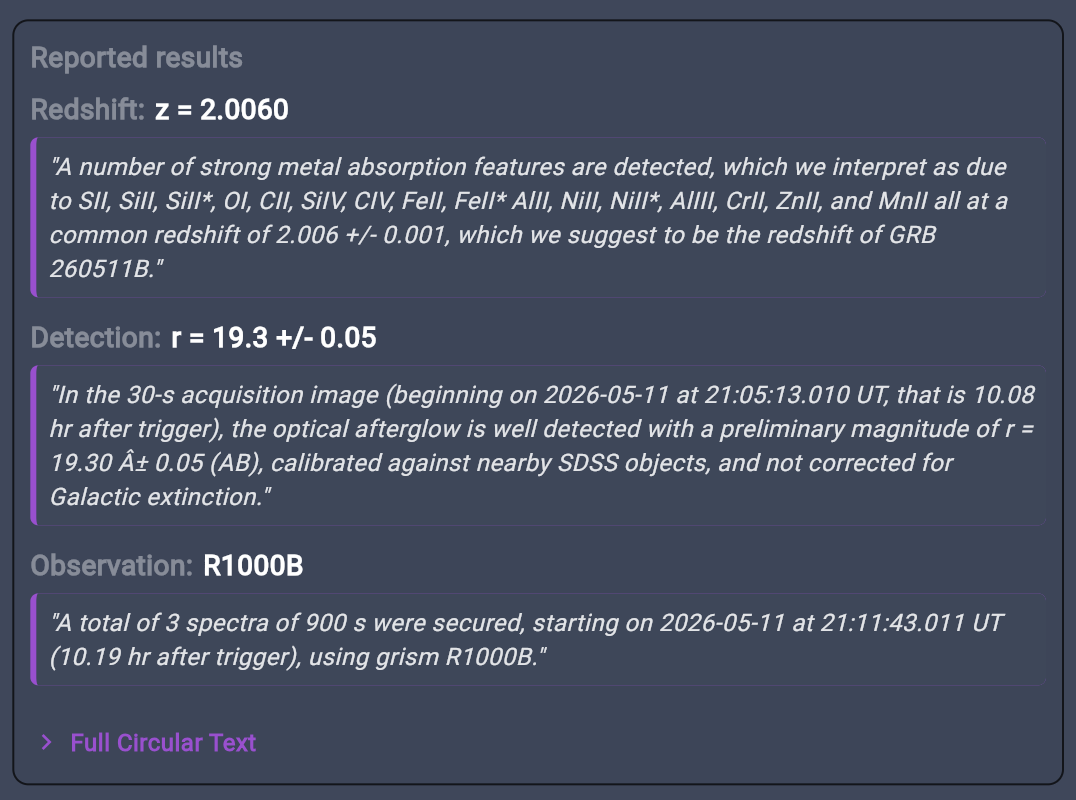}\\[-0.2em]
{\small (b) Extracted values and verification context (GCN Circular 44565)}
\end{minipage}
\caption{Report-level presentation of extracted Circular information in the \ac{} frontend. Panel (a) shows an individual report summary card with selection state, observatory/instrument metadata, classification icons, submitter and author information, extracted summary text, and Circular identifier. Panel (b) shows the expanded report details, where structured values such as redshift, detection magnitude, and observing setup are displayed together with the exact Circular excerpts from which they were extracted; the full Circular text can be opened below for verification and complete context. Both panels show the same report, extracted from GCN Circular 44565 on GRB~260511B: panel (b) is the expansion of the card in panel (a), which carries the Circular identifier in the interface.}
\label{fig:frontend-report-detail}
\end{figure*}

The \ac{} user interface exposes the follow-up database inside the source-information panel in both the web and smartphone applications. For GRB-like events, the section is labeled as optical afterglow and follow-up; for other transients, it appears as a general follow-up section. The frontend loads the event summary, displays the context figure when available, and shows report groups. A first summary view is intentionally compact: each Circular or community report is represented by a headline-level card that provides the reporting facility, the Circular subject or report title, the extracted result statement, the relevant contacts, and a short natural-language summary. Expanding the card exposes the extracted values including the observatory, instrument, subject, report identifiers, report URLs, contacts, submitters, authors, collaborations, extracted results, flags, redshift information, observations, stored Circular text, and the short summaries together with the source excerpts from which they were derived. The cards also provide access to the full Circular text for verification and broader context. Figure~\ref{fig:frontend-report-detail} illustrates the compact report card and the expanded view of the extracted values and their provenance.

The report cards use a common set of classifications (observation report, detection, upper limit, spectrum, and redshift) both as visual indicators and as filters. Active icons are highlighted while inactive icons are dimmed, and tooltips expose the corresponding category labels. An observer can therefore switch quickly between, for example, non-detections that constrain an afterglow, redshift or spectroscopy Circulars, and reports with positive detections. The filters operate on report groups rather than isolated photometry rows, preserving the surrounding context, contact information, and provenance. The same convention is used for automatically extracted GCN reports, direct observatory submissions, and report-management views.

The same panel provides the photometry download control. \ac{} users can request the CSV or VOTable exports described in Section~\ref{sec:data_products} through the API. The service returns a storage URL, and the frontend opens the file for download.

\subsection{Afterglow Light-Curve Fitting}
\begin{figure*}[!t]
\centering
\begin{minipage}{0.47\linewidth}
\centering
\includegraphics[width=0.98\linewidth]{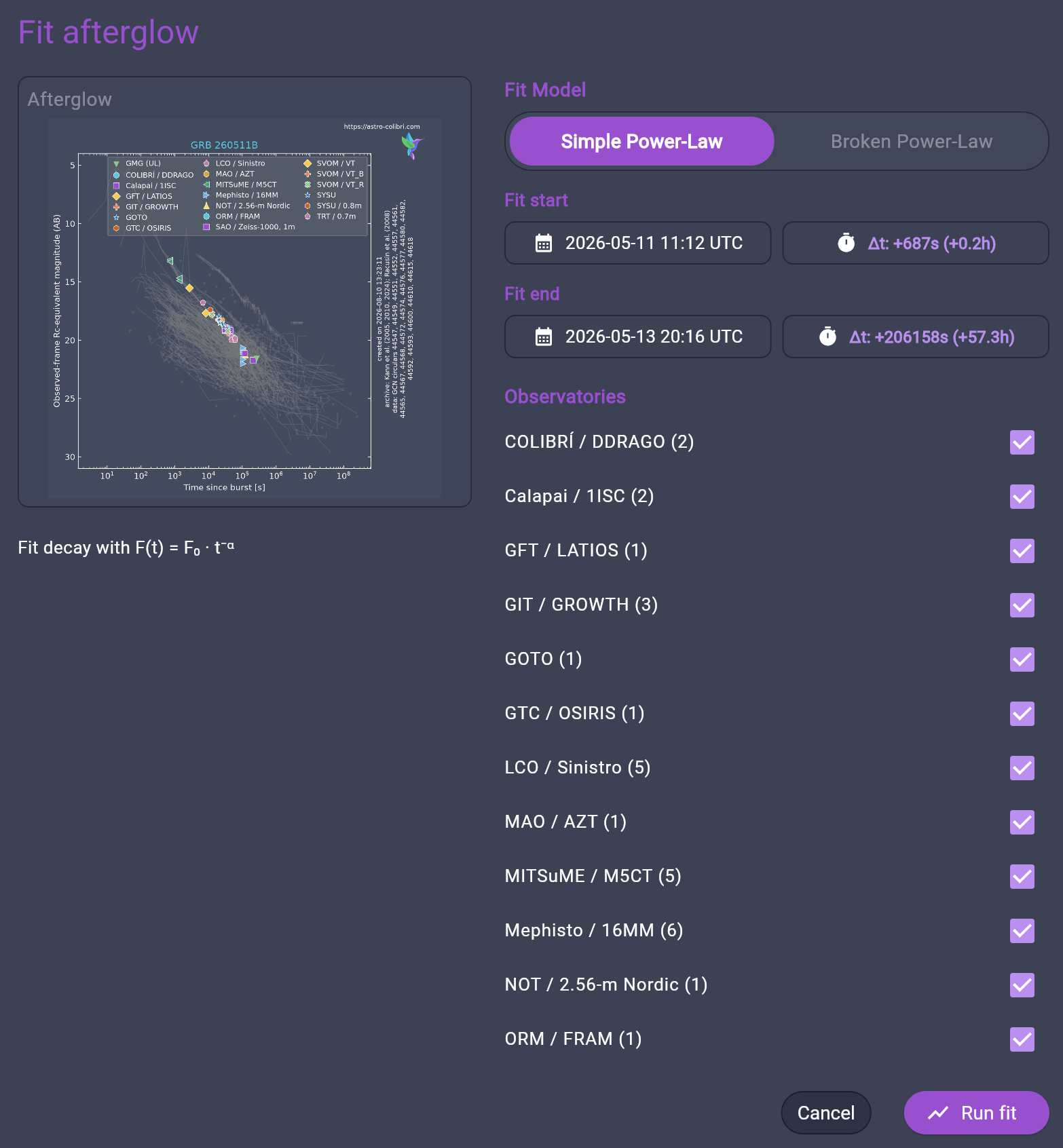}\\[-0.2em]
{\small (a) Fit-configuration dialog}
\end{minipage}\hfill
\begin{minipage}{0.51\linewidth}
\centering
\includegraphics[width=0.98\linewidth]{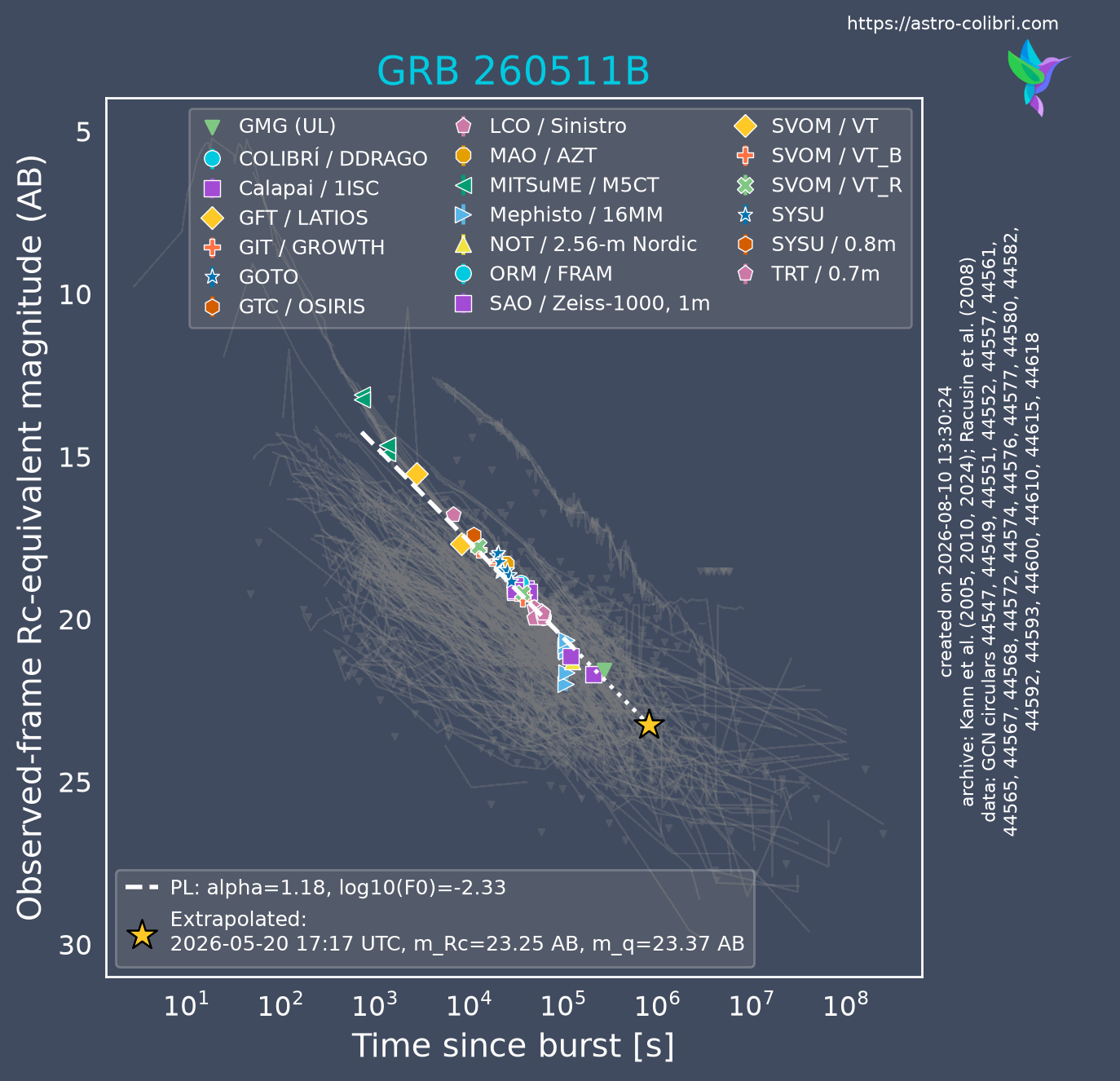}\\[-0.2em]
{\small (b) Generated fitted afterglow figure}
\end{minipage}
\caption{Frontend fitting workflow for the optical-afterglow product. Panel (a) shows the fit dialog, where a user selects the model family, time interval, and contributing observatory/instrument groups and may request an extrapolation to a future observing time. Panel (b) shows the resulting figure with the complete event photometry and archival comparison population, the curve fitted to the selected detections, its extension to the requested time, and the extrapolated point marked by a star.}
\label{fig:afterglow-fit-ui}
\end{figure*}

A light-curve fitting tool is integrated into the optical-afterglow panel to support rapid assessment and observation planning. The workflow, summarized in Figure~\ref{fig:afterglow-fit-ui}, is available to all authenticated \ac{} users. When the fit control is opened, the frontend first requests the event's fit metadata from the follow-up data endpoint. This response provides the number of usable photometric detections, the earliest and latest detection times as the default fit interval, the observatory/instrument groups and their numbers of measurements, and the current afterglow figure. The fit operates on the aggregated event-level record rather than on any single Circular: every measurement merged into the event from all of its reports is available for selection, and a fit repeated after further Circulars arrive will therefore use the enlarged data set.

As shown in Figure~\ref{fig:afterglow-fit-ui}(a), the fit interval can be edited either as UTC dates or as seconds since the event trigger, and individual observatory/instrument groups can be included or excluded. The fit uses only measurements classified as detections that have a positive time since trigger and a valid \rcab{} magnitude. Upper limits are excluded from the optimization, although they remain visible in the resulting context figure. An optional extrapolation control estimates the magnitude at a user-selected UTC time. When observatory visibility information is available, the interface initializes this time to the beginning of the relevant visibility window; otherwise, it retains a default future time that the user can override. The corresponding visibility plot can also be inspected directly from the panel.

Two temporal models are available. Time $t$ is measured in seconds since the event trigger, and $F_0$ is a relative-flux normalization. The simple model is
\begin{equation}
    F(t) = F_0\,t^{-\alpha},
\end{equation}
while the smoothly broken power law follows the commonly used Beuermann parametrization \citep{1999A&A...352L..26B}:
\begin{equation}
    F(t) = F_0\left[\left(\frac{t}{t_b}\right)^{\alpha_1 n}
    + \left(\frac{t}{t_b}\right)^{\alpha_2 n}\right]^{-1/n},
\end{equation}
where $t_b$ is the break time, $\alpha_1$ and $\alpha_2$ are the asymptotic temporal indices before and after the break, and $n$ controls the smoothness of the transition. At least two selected detections are required for the simple power law and at least five for the five-parameter broken-power-law model. These are implementation minima: five points leave no nominal degrees of freedom for the broken model, so substantially more measurements are required for meaningful parameter constraints. The fit dialog states this at the point of use, warning whenever the selected detections leave two or fewer degrees of freedom for the chosen model. The warning is advisory rather than blocking, since a weakly constrained fit can still serve as a quick look provided it is not mistaken for a constrained measurement. The backend converts each selected magnitude $m$ to relative flux, $F=10^{-0.4m}$, and performs a bounded, weighted nonlinear least-squares fit in flux space. Symmetric or asymmetric magnitude uncertainties are propagated to the corresponding asymmetric flux uncertainties, and each residual is weighted by the uncertainty on the side facing the model prediction. If no uncertainty is reported for a point, the current implementation adopts $0.1$~mag for the fit weighting. Local parameter uncertainties are returned when a covariance estimate can be obtained from the fit Jacobian. These uncertainties are statistical only. They propagate the reported photometric errors through the fit, and do not include the systematic component introduced by the common-scale conversion, in particular the assumed spectral index $\beta=0.7$ used where no measured value is available. Since the Circulars do not report an uncertainty on that index, a defensible systematic term cannot be attached to individual points, and inflating the errors by an assumed value would give the fitted parameters a false appearance of accounting for it. Users comparing fitted decay indices across events should therefore treat the quoted uncertainties as lower bounds.

The fit request returns the model parameters, the identities of the measurements and observatory groups used, an optional extrapolated \rcab{} magnitude, and the URL of a newly generated fit figure. The frontend presents the decay index and relative-flux normalization for a simple power law; for a broken power law, it additionally presents the two indices, break time, and smoothness parameter. If extrapolation was requested, the predicted magnitude and UTC time are highlighted. The generated plot in Figure~\ref{fig:afterglow-fit-ui}(b) retains the complete event photometry and archival comparison population, overlays the fitted curve across the selected interval, draws any extension to the extrapolation time separately, and marks the predicted point with a star. Users can reopen the dialog to change the selection and refit. These fits are intended as rapid descriptive and planning tools and inherit the photometric-normalization assumptions and data-quality caveats discussed above; they should not be interpreted as a substitute for publication-level physical afterglow modeling.

Even so, an early estimate of the temporal decay slope, together with the identification of a possible break, already carries physical information. It indicates whether the light curve behaves as expected for synchrotron emission from a decelerating blast wave \citep{1998ApJ...497L..17S}, and when combined with spectral information it constrains the characteristic frequencies $\nu_m$ and $\nu_c$ through the closure relations between the temporal and spectral indices \citep{2013NewAR..57..141G}. A decay index that departs from the standard expectations, or a break appearing earlier than anticipated, is exactly the kind of signal that justifies committing scarce resources: triggering radio observations, extending late-time monitoring, or requesting spectroscopy while the source is still bright enough. Read this way the fit is not a measurement to be published but a triage tool, indicating which events merit deeper investment during the hours in which that decision has to be made.

\subsection{Observer Contacts and Coordination}\label{sec:contacts}

Rapid access to the teams involved in a transient campaign is an important part of the follow-up overview. Contact records associated with GCN Circulars are assembled from header metadata, email addresses found in the body, and schema-constrained model output, while direct \ac{} observation reports retain the contact information supplied by their submitters. These contacts remain linked to their originating reports so that users can identify not only an email address, but also the corresponding observatory, instrument, Circular, and reported result.

Users may select individual report groups or all reports associated with an event and ask the frontend to prepare a contact email. The interface collects and deduplicates valid addresses from the report contacts and submitters. It then creates a message addressed to these recipients, with the transient name in the subject and a body listing the selected Circular identifiers, observatory/instrument combinations, and report subjects. The message is passed to the user's local email application through a pre-filled email link; \ac{} itself does not send the email or act as an intermediary for the subsequent correspondence.

Automated extraction can occasionally recover unattended addresses such as \code{noreply} or \code{do-not-reply} mailboxes. Before preparing the message, the interface identifies such addresses and displays a warning containing the associated observatory and links to the originating Circulars. These addresses are excluded unless the user explicitly chooses to retain them, and the dialog indicates whether another usable address is available for the same report. This keeps recipient selection under user control while preserving the provenance needed to verify questionable contact information. The objective is not automated mass communication, but to reduce the practical friction involved in confirming reported measurements, coordinating additional observations, requesting data, and initiating joint analyses or publications.

\begin{figure*}[!th]
\centering
\begin{nolinenumbers}
\resizebox{0.99\linewidth}{!}{%
\begin{tikzpicture}[
    node distance = 0.6cm and 1.0cm,
    box/.style = {
        rectangle,
        rounded corners=4pt,
        minimum width=3.8cm,
        minimum height=1.3cm,
        align=center,
        font=\sffamily\small,
        drop shadow={shadow xshift=1.5pt, shadow yshift=-1.5pt, fill=black!20}
    },
    hub/.style = {box, fill=hubgreen, draw=hubtext, text=hubtext, line width=1pt, font=\sffamily\bfseries\small},
    action/.style = {box, fill=actionblue, draw=actiontext, text=actiontext, line width=0.8pt, minimum height=1.9cm},
    synth/.style = {box, fill=synthgold, draw=synthtext, text=synthtext, line width=0.8pt, minimum height=1.9cm},
    arrow/.style = {-{Stealth[scale=1.2]}, line width=1pt, draw=gray!80}
]

    % --- Central Enabler ---
    \node (hub) [hub, minimum width=4.2cm] {Event-Level \\ Follow-Up View \\ \mdseries\footnotesize (Visual Overview, \\ \mdseries\footnotesize{Provenance \& Contacts})};

    % Each use case names the concrete product that delivers it, with a pointer
    % to the figure or section where that product is shown.
    % --- Upper Path: Immediate Real-Time Actions ---
    \node (coord) [action, above right=of hub, xshift=0.5cm] {\textbf{Real-Time Coordination} \\ \footnotesize Prioritize facilities \& \\ avoid duplicated efforts \\ \itshape via report cards \& \\ classification filters (Fig.~\ref{fig:frontend-report-detail})};
    \node (assess) [action, right=of coord] {\textbf{Coverage Assessment} \\ \footnotesize Track filters, epochs, \\ limits, specs \& redshifts \\ \itshape via per-observation rows \\ \& event summary (Sec.~\ref{sec:data_products})};
    \node (interp) [action, right=of assess] {\textbf{Rapid Interpretation} \\ \footnotesize Evaluate afterglow \\ brightness in context \\ \itshape via context figure \& \\ light-curve fit (Figs.~\ref{fig:afterglow-example},~\ref{fig:afterglow-fit-ui})};

    % --- Lower Path: Active Synthesis & Verification ---
    \node (collab) [synth, below right=of hub, xshift=0.5cm] {\textbf{Collaboration Building} \\ \footnotesize Identify and contact \\ active observing teams \\ \itshape via deduplicated contacts \\ \& pre-filled email (Sec.~\ref{sec:contacts})};
    \node (verify) [synth, right=of collab] {\textbf{Data Verification} \\ \footnotesize Trace metrics back to \\ Circulars \& exact excerpts \\ \itshape via source excerpts \& \\ stored full text (Fig.~\ref{fig:frontend-report-detail}b)};
    \node (offline) [synth, right=of verify] {\textbf{Offline Analyses} \\ \footnotesize Data sharing, VOTable \\ exports \& paper planning \\ \itshape via CSV/VOTable exports \\ (Sec.~\ref{sec:data_products})};

    % --- Final Endpoint ---
    \node (science) [hub, right=1.4cm of interp, yshift=-1.95cm, minimum width=4.0cm, fill=hubgreen!40] {Coordinated \\ Multi-Instrument \\ Science};

    % --- Connections ---
    % Split from Hub
    \draw [arrow] (hub.east) -- ++(0.3,0) |- (coord.west) node[pos=0.35, left, xshift=-0.05cm, font=\sffamily\bfseries, text=actiontext] {enables};
    \draw [arrow] (hub.east) -- ++(0.3,0) |- (collab.west);

    % Upper Flow
    \draw [arrow] (coord.east) -- (assess.west);
    \draw [arrow] (assess.east) -- (interp.west);

    % Lower Flow
    \draw [arrow] (collab.east) -- (verify.west);
    \draw [arrow] (verify.east) -- (offline.west);

    % Merge to final science box
    \draw [arrow] (interp.east) -- ++(0.35,0) |- (science.north west) node[pos=0.25, right, font=\sffamily\bfseries, text=hubtext] {supports};
    \draw [arrow] (offline.east) -- ++(0.35,0) |- (science.south west);

    % --- Section Braces / Background Labels ---
    \node [above=0.2cm of assess, font=\sffamily\bfseries\color{actiontext}] {Immediate Real-Time Actions (Minutes to Hours)};
    \node [below=0.2cm of verify, font=\sffamily\bfseries\color{synthtext}] {Long-Term Synthesis \& Archiving (Days to Months)};

\end{tikzpicture}
}
\end{nolinenumbers}
\caption{Scientific use cases supported by the \ac{} follow-up data, each annotated with the concrete product that delivers it and a pointer to where that product is described. A single event-level overview helps observers coordinate in real time, assess wavelength and temporal coverage, interpret afterglow brightness in context, identify collaborators, verify extracted values against their provenance, and organize offline analyses and publications efficiently.}
\label{fig:scientific-use-cases}
\end{figure*}
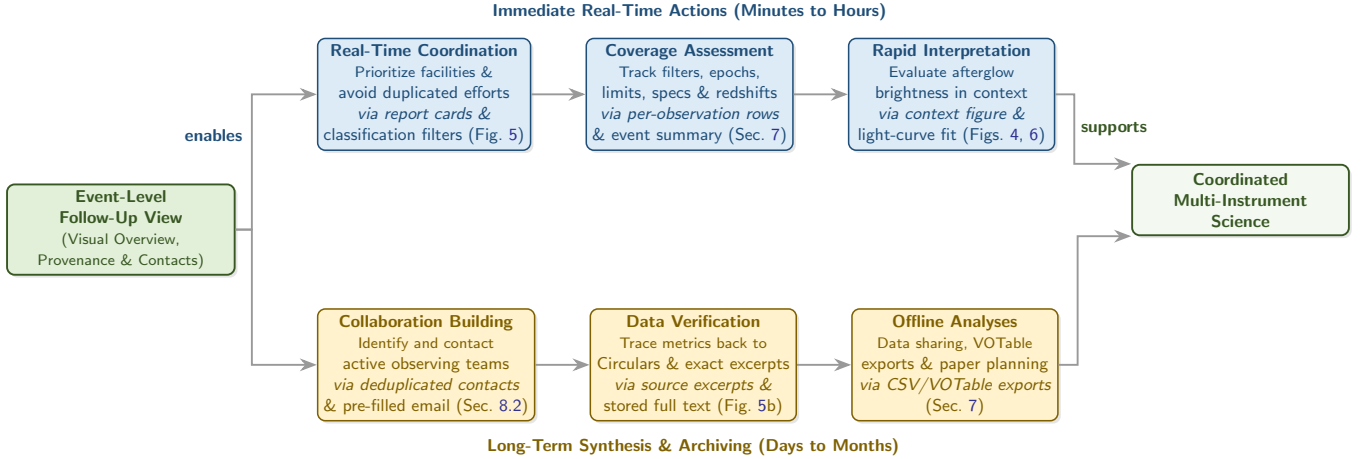

Figure~\ref{fig:scientific-use-cases} summarizes how these capabilities support both immediate follow-up coordination and longer-term scientific analysis.

\section{\ac{} Community Observation Reports}\label{sec:community_reports}
A complementary channel allows registered observatories to submit observation reports directly through the \ac{} web and mobile interfaces. This channel is intended for teams that have observed an active transient but are not yet ready to make the scientific results public through a GCN Circular. In many cases, data may still be proprietary, calibration may be incomplete, or the collaboration may prefer to coordinate interpretation before publishing magnitudes, spectra, classifications, or limits. A lightweight, informal \ac{} report lets these teams indicate that relevant observations exist, identify the responsible observatory or collaboration, and provide a contact point without requiring a full public announcement of the measurements themselves. This improves the discoverability of follow-up data during the active phase of a transient, makes it easier to include contributing teams in later offline analyses and publications, and provides a complementary coordination channel before a formal public announcement.

The report dialog is available to observatory members and administrators. A minimal report contains the event, observatory, contact name, and contact email, which is enough to indicate that the observatory has data on the source. Users may optionally add observation rows with time, filter or frequency/energy/messenger label, magnitude or flux, uncertainty, upper-limit state, redshift, comments, and a flag requesting regeneration of the optical-afterglow figure. These interface-based submissions are ingested by internal backend components, which enrich the observations in the same way as GCN-extracted rows, send observatory notifications, and, for GRB-like events with sufficient photometry, update the afterglow figure. As a result, public Circular-derived information and direct community reports appear together in the follow-up overview and exports.

Observation reports can also be created automatically by interfaces to partner follow-up programs. For example, the H.E.S.S. GRB follow-ups reported in real time through the dedicated H.E.S.S. pages~\citep{hessGRBRealtime} are mirrored as \ac{} observation-report entries, so that very-high-energy observations are visible alongside optical, X-ray, and other follow-up activity.

Newly started campaigns by Kilonova-Catcher ({\emph KNC}), the citizen-science program of the Global Rapid Advanced Network Devoted to the Multi-messenger Addicts ({\emph GRANDMA}), are automatically announced to \ac{} and represented as observation reports \citep{kncWeb}. For the Black Hole Target and Observation Manager ({\emph BHTOM}), all targets being observed by the network are listed in the \ac{} follow-up view, independently of whether a target was created from an \ac{} submission or directly in BHTOM \citep{bhtomWeb,2025RMxAC..59..167M}. The {\it R\'eseau Amateurs Professionnels pour les Alertes Scientifiques} ({\emph RAPAS}), follows the same principle: targets are created within \ac{}, announced through the \ac{} discussion forum, and represented by an observation-report entry \citep{rapasWeb}. These automatically generated reports serve the same purpose as manual community reports: they expose active follow-up, preserve the responsible network and contact path, and help identify teams with relevant data for coordinated analyses and publications. Follow-up programs interested in setting up similar automated submissions are invited to contact the \ac{} team at \href{mailto:astro.colibri@gmail.com}{\texttt{astro.colibri@gmail.com}}.

\section{Scientific Use Cases and Caveats}

The system is designed for real-time situational awareness and collaboration support. It answers operational questions quickly: what is the latest state of the (afterglow) emission, what is the approximate optical-afterglow brightness compared with archival GRBs, which teams have reported observations, which filters and epochs are covered, whether a redshift has been measured, and whom to contact for confirmation or data access. This is valuable during the first minutes to days after a transient, when observing resources must be allocated and duplicated effort should be minimized. It is also useful after the immediate observing phase: the event-level follow-up view provides a first-glance overview of available data, identifies the teams involved in the detection and follow-up campaign, and offers a practical route to contact the relevant observers for coordinated offline analyses, data sharing, and publication planning.

At the same time, the database should not be treated as a verified, publication-quality photometry archive. Free-text Circulars can contain ambiguity in time expressions, table formatting, filter naming, and calibration references, while cross-filter conversion to a common \rcab{} magnitude requires an assumed spectral shape when no index is reported. The system therefore retains Circular identifiers, report URLs, source excerpts, full text, extraction warnings, and contact records so that users can audit and confirm values. The purpose of the NLP component is to reduce the latency between report publication and community awareness, not to replace expert judgment or formal data releases. Before publication-level analysis, users should compare extracted rows with the original GCN Circulars and contact the reporting teams for final calibrated data.

\section{Discussion and Future Work}

The present implementation demonstrates the operational feasibility of integrating schema-constrained LLM extraction into a live astronomical platform. The archive-scale completion statistics reported above do not, however, measure field-level scientific accuracy, and the two evaluations that do address it are limited in complementary ways: the human audit is internal and measures precision but not recall, while the redshift cross-check against three published compilations is genuinely independent but covers a single scalar field. Both limitations share a root cause --- the lack of an independently annotated corpus of Circulars against which a complete extraction could be scored --- and neither can be removed by inspecting the pipeline output more carefully. Moreover, model updates and provider behavior can change an extraction even when the prompt and schema remain fixed. Recording the model family, retaining the full source text and supporting excerpts, and separating deterministic transformations from model output improve auditability and reproducibility, but do not remove the need for an independent annotated evaluation set and continued human review.

\subsection{Single-source association assumption}\label{sec:single_source_limitation}
The parser's association model assumes that each Circular concerns a single astrophysical source. This assumption is valid for the overwhelming majority of Circulars and allows each extracted report to be assigned to one event-level source record. The plural event-name field can represent aliases for that source, and the database can link multiple alerts or refinements that refer to it, but neither mechanism represents distinct candidate counterparts. A rare corner case occurs when a Circular reports several mutually exclusive possible associations. For example, \href{https://gcn.nasa.gov/circulars/45095}{GCN Circular 45095} identifies a transient by the Wide-field X-ray Telescope (WXT) onboard the Einstein Probe satellite as a likely stellar flare but retains two stars within the Followup X-ray Telescopes (FXT) error circle, StKM~1-1292 and PM~J15587+2351E, as alternative possible origins, with a conditional X-ray luminosity reported for each. The current pipeline does not automatically represent this one-to-many, conditional association, so such reports require manual review. A future extension should distinguish the transient event from its candidate counterparts in different wavelengths and attach measurements or derived quantities to the appropriate conditional association.

\subsection{Future work}\label{sec:future_work}
A curated, independent set of historical Circulars should be annotated to quantify precision and recall for observation classification, time parsing, magnitude extraction, upper-limit recognition, filter normalization, contact extraction, and redshift identification, together with numerical errors for continuous quantities. The absence of such a resource is precisely why recall is not reported for the schema as a whole in this work; Section~\ref{sec:redshift_crosscheck} bounds it for the one field where an external reference exists. The annotated corpora that do exist for this text domain, \emph{TDAC} and \emph{astroECR} \citep{alkan-etal-2022-tdac, alkan-etal-2024-enriching}, label named entities, coreference chains, astrophysical relations, and normalized celestial-object names; they were not built to enumerate the observation-level records targeted here, and so cannot supply the required ground truth. Measuring recall means listing, ahead of any extraction, every observation a Circular actually reports, together with its time, filter, exposure, and detection or limit status, and its attribution to the correct source. Producing that at a useful scale is a substantial annotation effort requiring domain expertise, which is why it has not been attempted for this task; it is nevertheless the only route to a defensible recall figure, and the same corpus would then serve as a fixed regression benchmark. Stored provider and model-version information can support such regression tests as LLM services evolve. Separately, the redshift cross-check of Section~\ref{sec:redshift_crosscheck} motivates extending the schema so that a reported value carries its own semantics (measurement, upper or lower limit, candidate-host photometric estimate, or conditional association) which would also address the one-to-many association case of Section~\ref{sec:single_source_limitation}. The direct-report channel also provides a possible route for human correction: observatories could confirm, edit, or supersede automatically extracted measurements from their own Circulars.

A natural next step is to connect the structured GRB-afterglow photometry assembled here to the recently launched, cloud-hosted Nuclear-physics and Multi-Messenger Astrophysics (NMMA; \citealt{2023NatCo..14.8352P}) modeling service integrated into \ac{}~\citep{NMMA_forthcoming}\footnote{\url{https://nmma.live}}. The service already supports on-demand Bayesian modeling and interpretation of photometric light curves using supernova models and provides kilonova models for gravitational-wave follow-up. Because the underlying NMMA framework also includes GRB-afterglow models, a future extension could pass the multi-band detections and upper limits extracted from GCN Circulars directly to these models. This would enable rapid physical interpretation and model comparison within \ac{}, without requiring users to install or operate the inference software locally.

The same extraction framework can be extended to other human-written rapid-report channels, such as The Astronomer's Telegram \citep{astronomersTelegram} and TNS AstroNotes \citep{tnsAstroNotes}. Reports could also be used to propose new \ac{} event entries when they describe a source absent from the database. Finally, the architecture can be expanded beyond optical/NIR GRB afterglows to other transient classes and messenger channels, provided that the schema and plotting conventions are adapted to the relevant data products.

\section{Software and Data Availability}\label{sec:software_availability}

The source code for the parsing and normalization pipeline is publicly available under the BSD 3-Clause License as \parser{} at \url{https://github.com/astro-transients/astro_colibri_circular_parser}. The version described in this paper is release v1.0.0, archived on Zenodo at \url{https://doi.org/10.5281/zenodo.21887735} \citep{circularParser}; the concept DOI \url{https://doi.org/10.5281/zenodo.21887734} always resolves to the most recent archived version. The repository provides installation and configuration instructions, a Python API and command-line interface, an example notebook with cached inputs for offline inspection, and a credential-free test suite. Live extraction with the included OpenAI implementation requires provider credentials; callers may instead supply another implementation of the extraction interface. Event association can use the public \ac{} API, a caller-supplied lookup backend, or be disabled for offline processing. The released package reproduces the Circular-to-payload path described above, while the persistent database, notification mechanisms, and web and mobile interfaces remain components of the \ac{} platform.

The source GCN Circulars are available from the public GCN archive \citep{gcnCirculars}. Derived report summaries and generated context figures can be viewed through \ac{}, whereas detailed CSV/VOTable exports and light-curve fitting require a user account. The event-level follow-up database is updated continuously and should be treated as a dynamic operational product rather than a versioned archival data release. Because it changes as new Circulars arrive, the aggregates quoted in Sections~\ref{sec:human_verification}--\ref{sec:redshift_crosscheck} are frozen as dated snapshots with checksums, and the scripts that produce them and the figures from those snapshots are distributed with the manuscript source.

\section{Conclusion}

We have described a new \ac{} follow-up component that converts free-text GCN Circulars into structured, event-linked observation reports. The system combines real-time Circular ingestion, event association, deterministic hints, schema-constrained LLM extraction, normalization, consistency checks, API storage, optical-afterglow figure generation, downloadable CSV/VOTable products, and a frontend that emphasizes provenance and contactability. By combining public Circular parsing with direct observatory submissions, \ac{} provides a unified view of transient follow-up campaigns and lowers the friction for coordinated multi-instrument science. Releasing the reusable Circular-to-payload pipeline as \parser{} makes the extraction and deterministic enrichment stages available for inspection, reproduction, and integration into other workflows. All 1,775 Circulars in the operational evaluation corpus from the first half of 2026 completed the parsing and automated-verification workflow, and an internal audit of 210 Circulars found 223 of its 231 assessed reports (96.5\%) fully confirmed, with the errors it did find localized to observation timing and facility attribution. An independent cross-check of the extracted redshifts against three published compilations found no misread value. Applied to the archive from 2016 onward, the pipeline has produced 68,393 observations from 26,811 reports across 5,787 events, of which 95.0\% of the optical measurements reach the common plotting scale. Quantifying recall across the full schema remains the main outstanding validation step, and awaits an independently annotated corpus of Circulars, which does not yet exist for this task; for redshift, where published compilations supply an external reference, the extraction recovers of order 90 to 97\% of catalogued values.

\begin{acknowledgments}
The authors acknowledge ANR (French National Research Agency) for its support of the project ``Multi-messenger Observations of the Transient Sky (MOTS)'' no. ANR-22-CE31-0012. The authors also acknowledge support from the ACME project funded through the European Union's Horizon Europe Research and Innovation programme under Grant Agreement No. 101131928. The authors acknowledge the use of AI-assisted writing and language editing tools during manuscript preparation; all content, interpretations, and conclusions were reviewed and approved by the authors. We thank Pierre Zweigenbaum, Cyril Grouin, and Felix Grèzes for collaboration and discussions on various aspects of the use of AI/NLP for time-domain astrophysics.
\end{acknowledgments}

\software{\parser{} \citep{circularParser},
          Matplotlib \citep{2007CSE.....9...90H},
          NumPy \citep{2020Natur.585..357H},
          pandas \citep{2020zndo...3509134R},
          SciPy \citep{2020NatMe..17..261V}
          }

%% Bibliography inlined for submission: this document needs no
%% .bbl, .bib or .bst file to compile.


\begin{thebibliography}{}
\expandafter\ifx\csname natexlab\endcsname\relax\def\natexlab#1{#1}\fi
\providecommand{\url}[1]{\href{#1}{#1}}
\providecommand{\dodoi}[1]{doi:~\href{http://doi.org/#1}{\nolinkurl{#1}}}
\providecommand{\doeprint}[1]{\href{http://ascl.net/#1}{\nolinkurl{http://ascl.net/#1}}}
\providecommand{\doarXiv}[1]{\href{https://arxiv.org/abs/#1}{\nolinkurl{https://arxiv.org/abs/#1}}}

% type= phdthesis
\bibitem[{A.~K. Alkan(2024)Alkan}]{alkan2024thesis}
Alkan, A.~K. 2024, \bibinfo{title}{Natural Language Processing for Analyzing Messages of Astrophysical Observations,} PhD thesis, Universit{\'e} Paris-Saclay

% type= inproceedings
\bibitem[{A.~K. Alkan {et~al.}(2024)Alkan, Grezes, Grouin, Schussler, \& Zweigenbaum}]{alkan-etal-2024-enriching}
Alkan, A.~K., Grezes, F., Grouin, C., Schussler, F., \& Zweigenbaum, P. 2024, \bibinfo{title}{Enriching a Time-Domain Astrophysics Corpus with Named Entity, Coreference and Astrophysical Relationship Annotations,} in Proceedings of the 2024 Joint International Conference on Computational Linguistics, Language Resources and Evaluation (LREC-COLING 2024), ed. N.~Calzolari, M.-Y. Kan, V.~Hoste, A.~Lenci, S.~Sakti, \& N.~Xue (Torino, Italia: ELRA and ICCL), 6177--6188.
\newblock \url{https://aclanthology.org/2024.lrec-main.545/}

% type= inproceedings
\bibitem[{A.~K. Alkan {et~al.}(2022)Alkan, Grouin, Schussler, \& Zweigenbaum}]{alkan-etal-2022-tdac}
Alkan, A.~K., Grouin, C., Schussler, F., \& Zweigenbaum, P. 2022, \bibinfo{title}{{TDAC}, The First Corpus in Time-Domain Astrophysics: Analysis and First Experiments on Named Entity Recognition,} in Proceedings of the First Workshop on Information Extraction from Scientific Publications, ed. T.~Ghosal, S.~Blanco-Cuaresma, A.~Accomazzi, R.~M. Patton, F.~Grezes, \& T.~Allen (Online: Association for Computational Linguistics), 131--139, \dodoi{10.18653/v1/2022.wiesp-1.15}

% type= misc
\bibitem[{{Astro-COLIBRI}(2026{\natexlab{a}})}]{APIdoc}
{Astro-COLIBRI}. 2026{\natexlab{a}}, {Astro-COLIBRI} API Documentation, \url{https://astro-colibri.science/apidoc}

% type= misc
\bibitem[{{Astro-COLIBRI}(2026{\natexlab{b}})}]{astroColibriReleases}
{Astro-COLIBRI}. 2026{\natexlab{b}}, {Astro-COLIBRI} Release Notes, \url{https://astro-colibri.science/releases}

% type= inproceedings
\bibitem[{S.~D. Barthelmy(1998)Barthelmy}]{barthelmy1998gcn}
Barthelmy, S.~D. 1998, \bibinfo{title}{The {GRB} Coordinates Network,} in AIP Conference Proceedings, Vol. 428, Gamma-Ray Bursts, 4th Huntsville Symposium, 99--103

% type= article
\bibitem[{K. {Beuermann} {et~al.}(1999){Beuermann}, {Hessman}, {Reinsch}, {Nicklas}, {Vreeswijk}, {Galama}, {Rol}, {van Paradijs}, {Kouveliotou}, {Frontera}, {Masetti}, {Palazzi}, \& {Pian}}]{1999A&A...352L..26B}
{Beuermann}, K., {Hessman}, F.~V., {Reinsch}, K., {et~al.} 1999, \bibinfo{title}{{VLT observations of GRB 990510 and its environment},} \aap, 352, L26

% type= misc
\bibitem[{{BHTOM}(2026)}]{bhtomWeb}
{BHTOM}. 2026, Black Hole TOM, \url{https://bhtom.space}

% type= misc
\bibitem[{{CDS VizieR}(2026)}]{vizierKann}
{CDS VizieR}. 2026, {J/A+A/686/A56}: {GRB} Optical Afterglow Catalogue Associated with Kann et al. 2024, \url{https://cdsarc.cds.unistra.fr/viz-bin/cat/J/A+A/686/A56}

% type= misc
\bibitem[{P. {Coppin}(2026){Coppin}}]{grbweb}
{Coppin}, P. 2026, {GRBweb}: A Comprehensive Catalog of Gamma-Ray Bursts, \url{https://user-web.icecube.wisc.edu/~grbweb_public/}

% type= article
\bibitem[{H. {Gao} {et~al.}(2013){Gao}, {Lei}, {Zou}, {Wu}, \& {Zhang}}]{2013NewAR..57..141G}
{Gao}, H., {Lei}, W.-H., {Zou}, Y.-C., {Wu}, X.-F., \& {Zhang}, B. 2013, \bibinfo{title}{{A complete reference of the analytical synchrotron external shock models of gamma-ray bursts},} \nar, 57, 141, \dodoi{10.1016/j.newar.2013.10.001}

% type= misc
\bibitem[{J. {Greiner}(2026){Greiner}}]{greinerTable}
{Greiner}, J. 2026, {Table of all well-localized Gamma-Ray Bursts}, \url{https://www.mpe.mpg.de/~jcg/grbgen.html}

% type= article
\bibitem[{C.~R. {Harris} {et~al.}(2020){Harris}, {Millman}, {van der Walt}, {Gommers}, {Virtanen}, {Cournapeau}, {Wieser}, {Taylor}, {Berg}, {Smith}, {Kern}, {Picus}, {Hoyer}, {van Kerkwijk}, {Brett}, {Haldane}, {del R{\'i}o}, {Wiebe}, {Peterson}, {G{\'e}rard-Marchant}, {Sheppard}, {Reddy}, {Weckesser}, {Abbasi}, {Gohlke}, \& {Oliphant}}]{2020Natur.585..357H}
{Harris}, C.~R., {Millman}, K.~J., {van der Walt}, S.~J., {et~al.} 2020, \bibinfo{title}{{Array programming with NumPy},} Nature, 585, 357, \dodoi{10.1038/s41586-020-2649-2}

% type= misc
\bibitem[{{H.E.S.S. Collaboration}(2026)}]{hessGRBRealtime}
{H.E.S.S. Collaboration}. 2026, {H.E.S.S.} {GRB} Follow-up Observations, \url{https://grbhess.github.io/}

% type= article
\bibitem[{D.~W. {Hogg} {et~al.}(2002){Hogg}, {Baldry}, {Blanton}, \& {Eisenstein}}]{hogg2002}
{Hogg}, D.~W., {Baldry}, I.~K., {Blanton}, M.~R., \& {Eisenstein}, D.~J. 2002, \bibinfo{title}{{The K correction},} arXiv e-prints, astro, \dodoi{10.48550/arXiv.astro-ph/0210394}

% type= article
\bibitem[{J.~D. {Hunter}(2007){Hunter}}]{2007CSE.....9...90H}
{Hunter}, J.~D. 2007, \bibinfo{title}{{Matplotlib: A 2D Graphics Environment},} Computing in Science and Engineering, 9, 90, \dodoi{10.1109/MCSE.2007.55}

% type= misc
\bibitem[{{International Virtual Observatory Alliance}(2026)}]{ivoaVotable}
{International Virtual Observatory Alliance}. 2026, {VOTable} Format Definition, \url{https://www.ivoa.net/documents/VOTable/}

% type= article
\bibitem[{D.~A. {Kann} {et~al.}(2010){Kann}, {Klose}, {Zhang}, {et~al.}}]{2010ApJ...720.1513K}
{Kann}, D.~A., {Klose}, S., {Zhang}, B., {et~al.} 2010, \bibinfo{title}{{The Afterglows of Swift-era Gamma-Ray Bursts. I. Comparing pre-Swift and Swift-era Long/Soft (Type II) GRB Optical Afterglows},} \apj, 720, 1513, \dodoi{10.1088/0004-637X/720/2/1513}

% type= misc
\bibitem[{D.~A. {Kann} {et~al.}(2005){Kann}, {Zeh}, \& {Klose}}]{KannCatalog2005}
{Kann}, D.~A., {Zeh}, A., \& {Klose}, S. 2005, {A catalog of optical/near-infrared data on GRB afterglows in the pre-Swift era. I. Light curve information}, \dodoi{10.48550/arXiv.astro-ph/0509466}

% type= article
\bibitem[{D.~A. {Kann} {et~al.}(2024){Kann}, {White}, {Ghirlanda}, {Oates}, {Melandri}, {Jel{\'\i}nek}, {de Ugarte Postigo}, {Levan}, {Martin-Carrillo}, {Paek}, {Izzo}, {Blazek}, {Th{\"o}ne}, {Ag{\"u}{\'\i} Fern{\'a}ndez}, {Salvaterra}, {Tanvir}, {Chang}, {O'Brien}, {Rossi}, {Perley}, {Im}, {Malesani}, {Antonelli}, {Covino}, {Choi}, {D'Avanzo}, {D'Elia}, {Dichiara}, {Fausey}, {Fugazza}, {Gomboc}, {Gorski}, {Granot}, {Guidorzi}, {Hanlon}, {Hartmann}, {Hudec}, {Jun}, {Kim}, {Kim}, {Klose}, {Klu{\'z}niak}, {Kobayashi}, {Kouveliotou}, {Lidz}, {Marongiu}, {Martone}, {Meintjes}, {Mundell}, {Murphy}, {Nalewajko}, {Park}, {Sz{\'e}csi}, {Smith}, {Stecklum}, {Steele}, {{\v{S}}trobl}, {Sung}, {Updike}, {Urata}, \& {van der Horst}}]{2024A&A...686A..56K}
{Kann}, D.~A., {White}, N.~E., {Ghirlanda}, G., {et~al.} 2024, \bibinfo{title}{{Fires in the deep: The luminosity distribution of early-time gamma-ray-burst afterglows in light of the Gamow Explorer sensitivity requirements},} \aap, 686, A56, \dodoi{10.1051/0004-6361/202348159}

% type= article
\bibitem[{W. {Kiendrébéogo} {et~al.}(2026){Kiendrébéogo} {et~al.}}]{NMMA_forthcoming}
{Kiendrébéogo}, W., {et~al.} 2026, \bibinfo{title}{{NMMA - Astro-COLIBRI: An Automated Light-Curve Classification Service in the Multi-Survey Era},} in prep.

% type= misc
\bibitem[{{Kilonova Catcher}(2026)}]{kncWeb}
{Kilonova Catcher}. 2026, {Kilonova Catcher}, \url{https://kilonovacatcher.in2p3.fr/}

% type= inproceedings
\bibitem[{P.~J. {Mikolajczyk} {et~al.}(2025){Mikolajczyk}, {Zieli{\'n}ski}, {Wyrzykowski}, {Krawczyk}, \& {Kotysz}}]{2025RMxAC..59..167M}
{Mikolajczyk}, P.~J., {Zieli{\'n}ski}, P., {Wyrzykowski}, L., {Krawczyk}, A., \& {Kotysz}, K. 2025, \bibinfo{title}{{Black Hole TOM - an Automatic Tool for Photometric Time-Domain Data},} in Revista Mexicana de Astronomia y Astrofisica Conference Series, Vol.~59, Revista Mexicana de Astronomia y Astrofisica Conference Series, 167--172, \dodoi{10.22201/ia.14052059p.2025.59.26}

% type= misc
\bibitem[{{NASA Goddard Space Flight Center}(2026)}]{gcnCirculars}
{NASA Goddard Space Flight Center}. 2026, {GCN Circulars}, \url{https://gcn.nasa.gov/circulars}

% type= misc
\bibitem[{{NASA/GSFC Swift Science Center}(2026)}]{swiftGRBTable}
{NASA/GSFC Swift Science Center}. 2026, {Swift} {GRB} Table, \url{https://swift.gsfc.nasa.gov/archive/grb_table/}

% type= article
\bibitem[{P.~T.~H. {Pang} {et~al.}(2023){Pang}, {Dietrich}, {Coughlin}, {et~al.}}]{2023NatCo..14.8352P}
{Pang}, P. T.~H., {Dietrich}, T., {Coughlin}, M.~W., {et~al.} 2023, \bibinfo{title}{{An Updated Nuclear-physics and Multi-messenger Astrophysics Framework for Binary Neutron Star Mergers},} Nature Communications, 14, 8352, \dodoi{10.1038/s41467-023-43932-6}

% type= article
\bibitem[{J.~L. {Racusin} {et~al.}(2008){Racusin} {et~al.}}]{2008Natur.455..183R}
{Racusin}, J.~L., {et~al.} 2008, \bibinfo{title}{{Broadband observations of the naked-eye gamma-ray burst GRB 080319B},} \nat, 455, 183, \dodoi{10.1038/nature07270}

% type= misc
\bibitem[{{RAPAS}(2026)}]{rapasWeb}
{RAPAS}. 2026, {RAPAS}: R\'eseau Amateurs Professionnels pour les Alertes Scientifiques, \url{https://rapas.imcce.fr/}

% type= misc
\bibitem[{J. {Reback} {et~al.}(2020){Reback}, {McKinney}, {jbrockmendel}, {Van den Bossche}, {Augspurger}, {Cloud}, {et~al.}}]{2020zndo...3509134R}
{Reback}, J., {McKinney}, W., {jbrockmendel}, {et~al.} 2020, {pandas-dev/pandas: Pandas}, Zenodo \dodoi{10.5281/zenodo.3509134}

% type= article
\bibitem[{P. {Reichherzer} {et~al.}(2023){Reichherzer}, {Sch{\"u}ssler}, {Lefranc}, {Becker Tjus}, {Mourier}, \& {Alkan}}]{2023Galax..11...22R}
{Reichherzer}, P., {Sch{\"u}ssler}, F., {Lefranc}, V., {et~al.} 2023, \bibinfo{title}{{Astro-COLIBRI 2{\textemdash}An Advanced Platform for Real-Time Multi-Messenger Discoveries},} Galaxies, 11, 22, \dodoi{10.3390/galaxies11010022}

% type= article
\bibitem[{P. {Reichherzer} {et~al.}(2021){Reichherzer}, {Sch{\"u}ssler}, {Lefranc}, {Yusafzai}, {Alkan}, {Ashkar}, \& {Becker Tjus}}]{2021ApJS..256....5R}
{Reichherzer}, P., {Sch{\"u}ssler}, F., {Lefranc}, V., {et~al.} 2021, \bibinfo{title}{{Astro-COLIBRI-The COincidence LIBrary for Real-time Inquiry for Multimessenger Astrophysics},} \apjs, 256, 5, \dodoi{10.3847/1538-4365/ac1517}

% type= article
\bibitem[{R. {Sari} {et~al.}(1998){Sari}, {Piran}, \& {Narayan}}]{1998ApJ...497L..17S}
{Sari}, R., {Piran}, T., \& {Narayan}, R. 1998, \bibinfo{title}{{Spectra and Light Curves of Gamma-Ray Burst Afterglows},} \apjl, 497, L17, \dodoi{10.1086/311269}

% type= misc
\bibitem[{F. {Sch{\"u}ssler} \& {Astro-COLIBRI team}(2026){Sch{\"u}ssler} \& {Astro-COLIBRI team}}]{circularParser}
{Sch{\"u}ssler}, F., \& {Astro-COLIBRI team}. 2026, {astro-colibri-circular-parser} v1.0.0: {AI}-assisted parsing of {GCN} Circulars into structured follow-up observations, Zenodo, \url{https://doi.org/10.5281/zenodo.21887735}; source repository at \url{https://github.com/astro-transients/astro_colibri_circular_parser} \dodoi{10.5281/zenodo.21887735}

% type= article
\bibitem[{V. {Sharma} {et~al.}(2026){Sharma}, {Agarwala}, {Racusin}, {Singer}, {Barna}, {Burns}, {Coughlin}, {Dutko}, {Elliott}, {Gupta}, {Mahabal}, \& {Mukund}}]{2026ApJS..283...30S}
{Sharma}, V., {Agarwala}, R., {Racusin}, J.~L., {et~al.} 2026, \bibinfo{title}{{Large Language Model-driven Analysis of General Coordinates Network (GCN) Circulars},} \apjs, 283, 30, \dodoi{10.3847/1538-4365/ae2e9c}

% type= article
\bibitem[{V. {Sotnikov} \& A. {Chaikova}(2023){Sotnikov} \& {Chaikova}}]{2023Galax..11...63S}
{Sotnikov}, V., \& {Chaikova}, A. 2023, \bibinfo{title}{{Language Models for Multimessenger Astronomy},} Galaxies, 11, 63, \dodoi{10.3390/galaxies11030063}

% type= misc
\bibitem[{{The Astronomer's Telegram}(2026)}]{astronomersTelegram}
{The Astronomer's Telegram}. 2026, The Astronomer's Telegram, \url{https://www.astronomerstelegram.org/}

% type= misc
\bibitem[{{Transient Name Server}(2026)}]{tnsAstroNotes}
{Transient Name Server}. 2026, {TNS AstroNotes}, \url{https://www.wis-tns.org/astronotes}

% type= article
\bibitem[{P. {Virtanen} {et~al.}(2020){Virtanen}, {Gommers}, {Oliphant}, {Haberland}, {Reddy}, {Cournapeau}, {Burovski}, {Peterson}, {Weckesser}, {Bright}, {van der Walt}, {Brett}, {Wilson}, {Millman}, {Mayorov}, {Nelson}, {Jones}, {Kern}, {Larson}, {Carey}, {Polat}, {Feng}, {Moore}, {VanderPlas}, {Laxalde}, {Perktold}, {Cimrman}, {Henriksen}, {Quintero}, {Harris}, {Archibald}, {Ribeiro}, {Pedregosa}, {van Mulbregt}, \& {SciPy 1.0 Contributors}}]{2020NatMe..17..261V}
{Virtanen}, P., {Gommers}, R., {Oliphant}, T.~E., {et~al.} 2020, \bibinfo{title}{{SciPy 1.0: fundamental algorithms for scientific computing in Python},} Nature Methods, 17, 261, \dodoi{10.1038/s41592-019-0686-2}

\end{thebibliography}
\end{document}